\documentclass[10pt,a4paper,twocolumn,english,twocolumns,prb]{revtex4}
\usepackage[T1]{fontenc}
\usepackage[latin9]{inputenc}
\usepackage{float}
\usepackage{amssymb}
\usepackage{graphicx}

\makeatletter

\pdfpageheight\paperheight
\pdfpagewidth\paperwidth

\@ifundefined{textcolor}{}
{%
 \definecolor{BLACK}{gray}{0}
 \definecolor{WHITE}{gray}{1}
 \definecolor{RED}{rgb}{1,0,0}
 \definecolor{GREEN}{rgb}{0,1,0}
 \definecolor{BLUE}{rgb}{0,0,1}
 \definecolor{CYAN}{cmyk}{1,0,0,0}
 \definecolor{MAGENTA}{cmyk}{0,1,0,0}
 \definecolor{YELLOW}{cmyk}{0,0,1,0}
}

\renewcommand\[{\begin{equation}}
\renewcommand\]{\end{equation}}

\makeatother

\usepackage{babel}
\begin{document}

\title{Generation and Manipulation of Localized Modes in Floquet Topological
Insulators}

\author{Yaniv Tenenbaum Katan and Daniel Podolsky}

\affiliation{Physics Department, Technion -- Israel Institute of Technology, Haifa
32000, Israel}
\begin{abstract}
We investigate Floquet Topological Insulators in the presence of spatially-modulated
light. We extend on previous work to show that light can be used to
generate and control localized modes in the bulk of these systems.
We provide examples of bulk modes generated through modulation of
different properties of the light, such as its phase, polarization
and frequency. We show that these effects may be realized in a variety
of systems, including a zincblende  and honeycomb models, and also
provide a generalization of these results to three dimensional systems.

\end{abstract}
\maketitle

\section{Introduction}

Topological insulators have attracted great interest in recent years.
These materials are predicted to display many interesting effects,
such as fractionalized excitations and gapless boundary properties.
Yet, despite intense experimental efforts, only a handful of realizations
of intrinsic topological insulators are currently known \cite{KaneMele,Qi Zhang,Fu_Kane,Bernevig,Konig Wiedrmann,Xia Qian Hsieh,Hsieh Qian,Hassan_Kane,Kitagawa_Berg_Demler_Rudner_033429,Kitagawa_Berg_Demler_Rudner_235114}.
As a consequence, many recent proposals have focused on methods to
engineer systems with topological properties.

One such proposal that has attracted significant attention was suggested
by Lindner et. al. \cite{LRG}, who demonstrated that time periodic
perturbations can generate topological characteristics. This may be
achieved, for example, by shinning light on a conventional insulator.
These systems, named ``Floquet Topological Insulators'' (FTIs),
are predicted to display insulating behavior at the bulk that co-exists
with metallic conductivity at the edges. In addition, FTIs are predicted
to display many intriguing effects such as Dirac cones in three dimensions
\cite{Lindner3d} and Floquet Majorana fermions \cite{Jiang} in superconductors.
Proposals for FTIs include a wide range of solid state and atomic
realizations \cite{Fertig Gu Arovas Auerbach,Kitagawa_Oka_Brataas_Fu_Demler,LRG}.
The direct observation of protected edge modes in photonic crystals
\cite{Photonic Crystals 2,Photonic Crystal} has demonstrated that
these proposals have experimental realizations which may lead to future
practical applications.

In previous work \cite{Tenebaum Katan  D. Podolsky} we studied FTIs
when the light is not uniform in space. We found that spatial modulation
of the light can give rise to interesting effects in these so called
``modulated FTIs''. For example, we studied a zincblende  model
driven at resonance by linearly polarized light, and found that domain
walls and vortices in the phase of the light can give rise to localized
modes and fractionalized excitations in the bulk of this system. 

In this paper, we extend these results to much more general conditions
than previously considered. In addition to configurations involving
modulation in the phase of the light, we provide new schemes to induce
localized modes in the bulk of FTIs where the frequency and the polarization
of the radiating light vary in space. We establish these results for
systems that are driven both on and off resonance by the light. In
addition, we demonstrate that these effects not only apply to insulators
of the zincblende type but can be also be generalized to semi-metals
like graphene. Furthermore, we provide the first example of a three-dimensional
modulated FTI. These results demonstrate the versatility of modulated
FTIs, and may have practical applications in photonic crystals and
solid state devices. 

\begin{figure}[b]
\includegraphics[width=1\columnwidth]{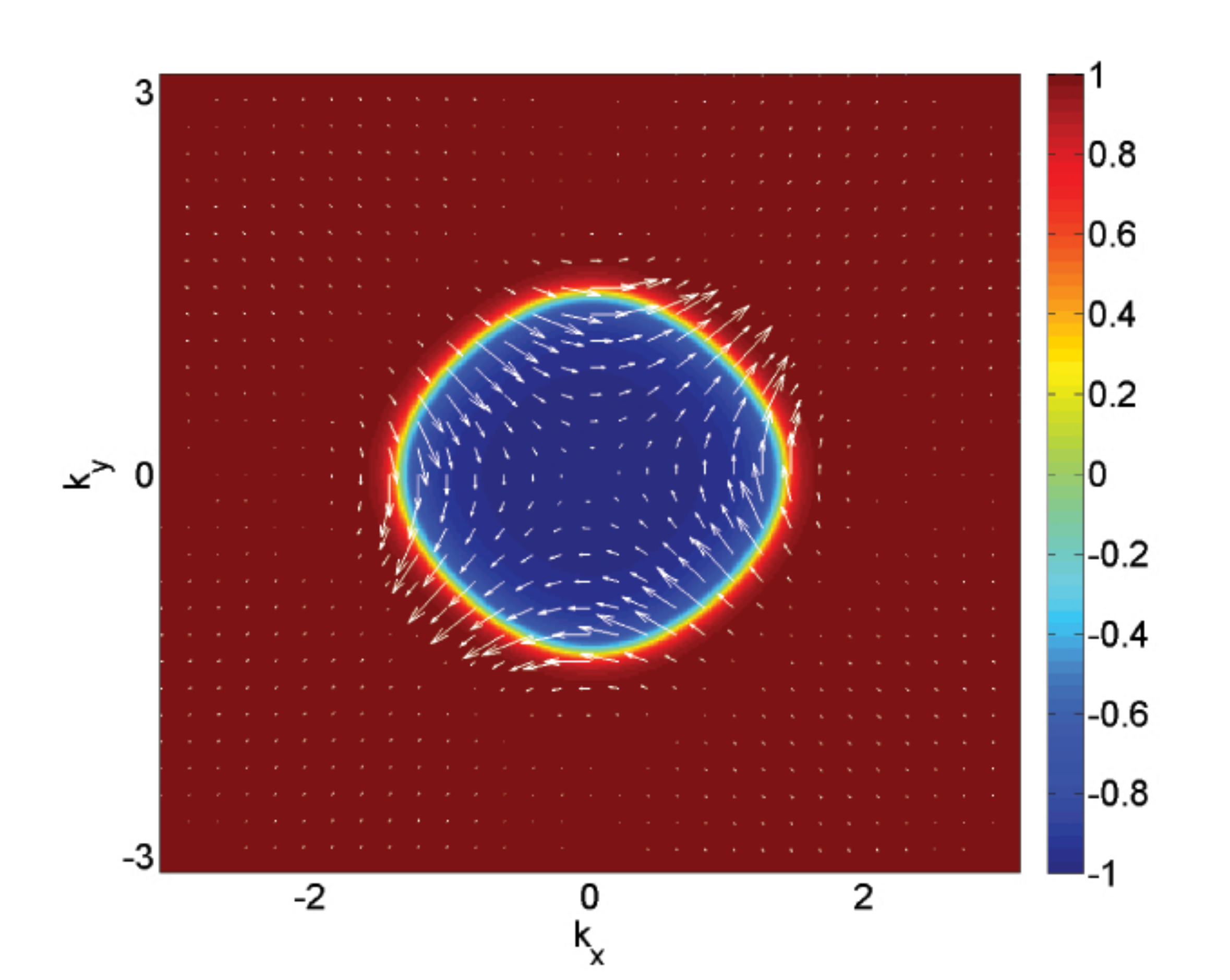}

\caption{$\hat{n}_{k}$, defined in Eq. (\ref{eq:Hf Definition}), for the
zincblende  model, Eq. (\ref{eq:HgTe d(k)}), with linearly polarized
light, $\alpha=0$. The arrows represent the $x$ and $y$ components
of $\hat{n}_{k}$ and the color map shows the $z$ component. Note
that $\hat{n}_{k}$ is in a hedgehog configuration, as it wraps the
unit sphere exactly once. This corresponds to $C_{F}=1$. \label{Flo:n(k)}}
\end{figure}

This article is organized as follows: In Sec. \ref{sec:General-Description}
we briefly review the concept of FTI and provide a short description
of the phenomenology of these phases. In Sec. \ref{sec:Square-lattice-model}
we summarize previous results for a square lattice zincblende model
irradiated by on-resonance light, and extend the analysis to light
with space dependent frequency. We also provide an analysis of the
effect of particle-hole symmetry breaking. In Sec. \ref{sub:Graphene}
we introduce modulated FTIs in graphene irradiated by light with on-resonance
and off-resonance frequencies. In particular, we study the effect
of light with space-dependent phase and polarization. In Sec. \ref{sub:3D-Topological-Insulator}
we provide an extension of modulated FTI to a three dimensional model
of a cubic lattice irradiated by on-resonance light with linear polarization.

\section{General Description\label{sec:General-Description}}

Let us begin by reviewing the concept of Floquet Topological Insulators
(FTI). For simplicity, consider a $2\times2$ Bloch Hamiltonian in
two dimensions

\begin{equation}
\tilde{H}_{k}=\vec{d}_{k}\cdot\vec{\sigma}+\epsilon_{k}I_{2\times2}\label{eq:H2x2}
\end{equation}
where $\vec{\sigma}$ denotes a vector of the Pauli matrices. Equation
(\ref{eq:H2x2}) has two energy bands for each $k$ value, with $E_{\pm}=\epsilon_{k}\pm\left|\vec{d}_{k}\right|$.
If $\vec{d_{k}}\neq0$ everywhere on the Brillouin zone, we can use
the TKNN formula \cite{Thouless} to define the topologically-invariant
Chern number of the occupied (lower energy) band \cite{Bernevig}

\begin{equation}
\begin{array}{c}
C=\frac{1}{4\pi}\int_{BZ}d^{2}k\hat{d}_{k}\cdot\left(\partial_{k_{x}}\hat{d}_{k}\times\partial_{k_{y}}\hat{d}_{k}\right)\end{array}\label{eq:C TKNN}
\end{equation}
Equation (\ref{eq:C TKNN}) counts the number of times $\hat{d}_{k}$
wraps around the unit sphere as $\vec{k}$ runs over the Brillouin
zone. Physically speaking, $C$ gives the net number of chiral modes
at the edge of the sample.

As shown in Ref. \cite{LRG}, even in cases where $C=0$, it is possible
to induce topological properties by perturbing the system in a time-periodic
fashion,
\[
H(t)=\tilde{H}_{k}+V(t),
\]
where $V(t+\tau)=V(t)$. The time evolution is then given by the Floquet
theorem \cite{Eastham}, which states that the solutions of the Schrödinger
equation can be written as $\psi\left(t\right)=\sum_{a}e^{i\varepsilon_{a}t}\varphi_{a}\left(t\right)$,
with $\varphi_{a}\left(t\right)=\varphi_{a}(t+\tau)$. The quasi-energies,
$\varepsilon_{a}$, are conserved quantities that are defined modulo
$\omega=\frac{2\pi}{\tau}$. They describe the evolution of states
over a full cycle.

The states $\varphi_{a}$ and their corresponding quasi-energies $\varepsilon_{a}$
satisfy the eigenvalue problem,

\[
H_{F}\varphi_{a}(t)=\varepsilon_{a}\varphi_{a}(t)
\]
where the ``Floquet Hamiltonian'' $H_{F}$ is defined as$C_{F}$
\[
e^{-iH_{F}\tau}\equiv U(t+\tau,t)
\]
and $U(t+\tau,t)$ is the time evolution operator over a full cycle,
\begin{equation}
U\left(t+\tau,t\right)=T\left\{ \exp\left(-i\int_{t}^{t+\tau}H\left(t'\right)dt'\right)\right\} .\label{eq:Udef}
\end{equation}
Here, $T$ is the time ordering operator. In this paper, we evaluate
the time evolution operator numerically by discretizing the time interval
$t\in\left[0,\tau\right]$, $\tau=\frac{2\pi}{\omega}$ and computing
the time-ordered product of $e^{-iH(t)\Delta t}$ over the sub-intervals
$\Delta t$. We then extract the Floquet Hamiltonian by computing
the logarithm of the operator $U$.

For two level systems, the Floquet Hamiltonian can be written most
generally as

\begin{equation}
H_{F}=\vec{n}_{k}\cdot\vec{\sigma}+\epsilon_{k}I_{2\times2}.\label{eq:Hf Definition}
\end{equation}
Provided that $\vec{n}_{k}$ does not vanish over the Brillouin zone,
we can define a new topological invariant $C_{F}$ associated with
$H_{F}$,
\begin{equation}
\begin{array}{c}
C_{F}=\frac{1}{4\pi}\int_{BZ}d^{2}k\hat{n}_{k}\cdot\left(\partial_{k_{x}}\hat{n}_{k}\times\partial_{k_{y}}\hat{n}_{k}\right)\end{array}\label{eq:CF}
\end{equation}
A Floquet topological insulator is characterized by non-vanishing
$C_{F}$ \cite{LRG}. Similarly to time-independent topological insulators,
a non-zero value of $C_{F}$ implies the existence of topologically-protected
chiral states at the edge of the sample. 

We note that, whereas $C_{F}\ne0$ is a sufficient condition for a
system to be topological, it is not necessary in the case of Floquet
insulators. The value of $C_{F}$ assigned to a given band counts
the difference between the number of right and left moving modes in
a given edge above the band, \emph{minus} the difference between right
and left moving modes in the same edge below that band. In the Floquet
spectrum, quasi-energy is periodic. Then, for example, it is possible
for every band in the system to have a right moving mode both above
and below itself. In this situation, every band has $C_{F}=0$, but
the system is nevertheless topological. The full characterization
of edge states in two-dimensional FTIs requires the introduction of
an extra topological invariant in addition to $C_{F}$, as discussed
in detail in Ref. \cite{Rudner}, where the new invariant is called
$W$. In this paper we do not consider situations where only $W$
is modulated in space (and $C_{F}$ is not), although it would be
interesting to consider such situations.

\section{Zincblende Model\label{sec:Square-lattice-model}}

\subsection{Previous results}

In this section we give an explicit example of an FTI. The results
presented here can be found in Refs. \cite{LRG,Tenebaum Katan  D. Podolsky}.

Consider the Hamiltonian

\begin{equation}
\begin{array}{c}
H_{k}=\left(\begin{array}{cc}
\tilde{H}_{k} & 0\\
0 & \tilde{H}_{-k}^{*}
\end{array}\right).\end{array}\label{eq:H 4x4}
\end{equation}
where $\tilde{H}_{k}$ is given by Eq. (\ref{eq:H2x2}), with

\begin{equation}
\begin{array}{c}
\vec{d}_{k}=\left(A\sin k_{x},A\sin k_{y},M+2B\left(\cos k_{x}+\cos k_{y}-2\right)\right)\end{array}\label{eq:HgTe d(k)}
\end{equation}
and $A,B,M$ are constants. This model can describe, for example,
HgTe/CdTe quantum wells. In this case $\tilde{H}_{k}$ $\left[\left(\tilde{H}_{-k}^{*}\right)\right]$
acts on the subspace spanned by the $J_{z}=\left(\frac{1}{2},\frac{3}{2}\right)$
$\left[J_{z}=\left(-\frac{1}{2},-\frac{3}{2}\right)\right]$ states
respectively and $\vec{d}_{k}$ describes the dispersion of the bands
including the effects of spin-orbit interaction. 

Note that $\tilde{H}_{k}$ and $\tilde{H}_{-k}^{*}$ are related by
a time reversal (TR) transformation, such that Eq. (\ref{eq:H 4x4})
is TR invariant. This implies that the overall Chern number is zero.
However, it is still possible to obtain a TR protected topological
phase in which the Chern number assigned to the $2\times2$ block
$\tilde{H}_{k}$ is nonzero. Explicit calculation yields that $C=\frac{1}{2}\left(1+sign\left(\frac{B}{M}\right)\right)$.
We work in the parameter space $M>0$, $B<0$, for which $\tilde{H}_{k}$
is trivial, and we add a time-periodic potential in order to induce
the topology.

As the time dependent potential, we use perturbations that do not
connect the two Hamiltonian blocks and perform the analysis for the
$2\times2$ block $\tilde{H}_{k}$. We choose 

\begin{equation}
\begin{array}{c}
H(t)=\vec{d}_{k}\cdot\vec{\sigma}+\epsilon_{k}I_{2\times2}+\vec{V}_{k}\cdot\vec{\sigma}\cos\left(\omega t+\alpha\right)\end{array}\label{eq:Perturbed Hamiltonian}
\end{equation}
where $\alpha$ is the delay phase of the external perturbation and
$\omega$ is its frequency. For simplicity, we take $\vec{V}_{k}=V_{0}\hat{z}$
in what follows. Equation (\ref{eq:Perturbed Hamiltonian}) can describe,
for example, the effect of linearly polarized light in HgTe/CdTe quantum
wells \cite{LRG}.

Let us first examine the effect of spatially uniform light. We consider
light that whose frequency $\omega$ is on-resonance, such that it
connects states in the valence and conduction bands directly. We use
periodic boundary conditions and compute the topological invariant
$C_{F}$ through direct calculation of $\hat{n}_{k}$. Figure \ref{Flo:n(k)}
shows that for $\omega>M+4|B|$ $\hat{n}_{k}$ wraps once around the
unit sphere as $k$ runs over the Brillouin zone, in correspondence
with $C_{F}=1$.\cite{LRG} The topological properties of the radiated
system has strong dependence on the details on the light. For example,
we find that for $\omega=M+|B|$, $C_{F}=-2$ with two localized states
at each edge of the system. Of these, one corresponds to zero quasi
energy while the other corresponds to $\frac{\omega}{2}$.

As shown in Ref. \cite{Tenebaum Katan  D. Podolsky}, there is a close
analogy between the Floquet description of this system and the Hamiltonian
of a $p_{x}+ip_{y}$ superconductor (pSC)\cite{Schrieffer}. This
analogy is exact provided $\epsilon_{k}=0$, in which case the system
is invariant under the particle-hole (PH) transformation which exchanges
the valence and conduction bands. Then, in the rotating wave approximation
(RWA), the Floquet Hamiltonian at low energies becomes 
\begin{equation}
H_{F}\approx\left(\begin{array}{cc}
\zeta_{k} & \Delta_{0}e^{-i\alpha}\left(k_{x}-ik_{y}\right)\\
\Delta_{0}e^{i\alpha}\left(k_{x}+ik_{y}\right) & -\zeta_{k}
\end{array}\right)\label{eq:P-SC Analogue Hf}
\end{equation}
where $\zeta_{k}=\frac{A^{2}}{2M}k^{2}-\mu$, $\mu=\frac{\omega}{2}-M$,
and $\Delta_{0}=\frac{V_{0}A}{2M}$. This effective Hamiltonian resembles
that of a superconductor in a Nambu-Gorkov form. The analogy can be
seen graphically in Fig$.$ \ref{Flo:n(k)}, in which the superconducting
order parameter is seen to have a $p_{x}+ip_{y}$ symmetry. The analogy
is imperfect; while in an actual superconductor the Nambu basis describes
particle and hole states, Eq$.$ (\ref{eq:P-SC Analogue Hf}) acts
on two particle-like states corresponding to valence and conduction
bands of the Floquet problem. Hence, the spectrum of Eq$.$ (\ref{eq:P-SC Analogue Hf})
will match that of the corresponding pSC, but the Hilbert spaces are
different. The nature of the states in the two cases is related by
a particle-hole transformation. 

Notice that in Eq. (\ref{eq:P-SC Analogue Hf}), the delay phase $\alpha$
plays the role of the superconducting phase. Correspondingly, space
modulation of $\alpha$ leads to many intriguing effects. For example,
in a domain wall configuration, in which the phase $\alpha$shifts
by $\pi$at $y=0$, localized modes with zero quasi energy appear
in the vicinity of the domain wall. Similarly, in a vortex configuration,
in which the phase $\alpha$ winds by $2\pi$ about a point, a state
with zero quasi-energy and fractional charge is localized at the vortex
core. The quasi-stationary modes are analogous to the well known zero
modes of $\pi$ junctions and pSC vortices \cite{TeoKane,Read &Green}. 

The above results can be reproduced with circularly polarized light,
which breaks TR explicitly. In this case, the light can be described
by a vector potential $\vec{A}\left(t\right)=A_{0}\left(\cos\left(\omega t+\alpha\right),\pm\sin\left(\omega t+\alpha\right),0\right)$,
where $+$$\left(-\right)$ denotes left-handed (right-handed) polarization.
$\vec{A}\left(t\right)$ is then implemented in the Hamiltonian by
the minimal substitution $\vec{k}\to\vec{k}+\vec{A}$. The resulting
low-energy Floquet Hamiltonian is identical to Eq. (\ref{eq:P-SC Analogue Hf}),
up to a constant shift in the initial phase of the light. We therefore
conclude that the presence of gapless modes is unaffected by breaking
of TRS, provided that the two Hamiltonian blocks in Eq. (\ref{eq:H 4x4})
do not mix.

\subsection{Position-Dependent frequency}

\begin{figure}
\includegraphics[width=0.55\columnwidth]{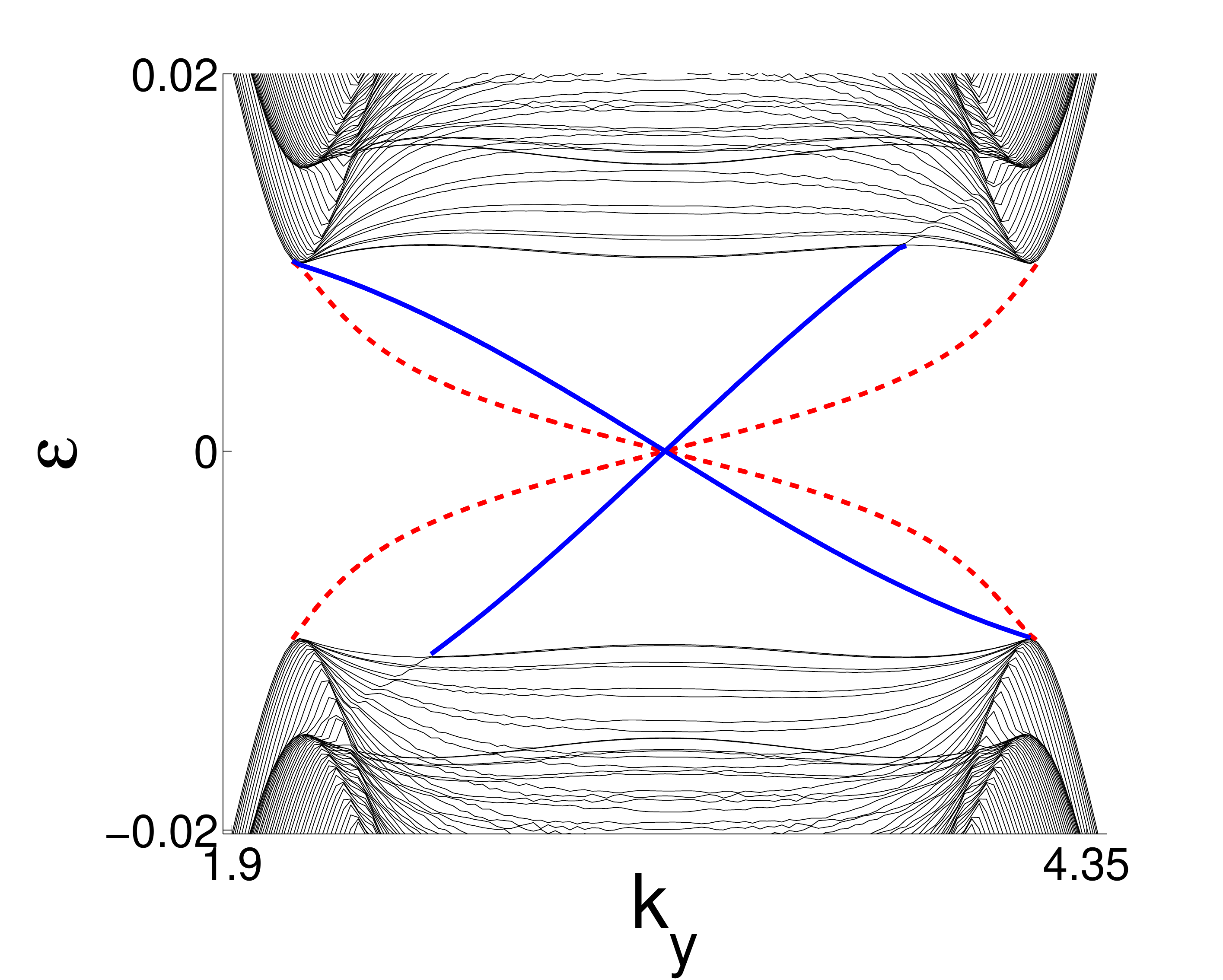}\caption{Floquet spectrum of the zincblende model irradiated by on-resonance
light with a space dependent frequency. The left half of the sample
has $\omega=2.7$ and the half has $\omega'=1.35$. The two solid
blue lines denote edge modes and the dashed red lines denote interface
modes. Note that the dispersion of the midgap states is not reflection
symmetric. Results are for $A=-B=0.05,M=1$ and $V_{0}=1$. \label{Flo: HgTe Lin with space dependent w}}
\end{figure}

As we now demonstrate, localized bulk modes can also be generated
by light whose frequency is varying in space. We first consider a
simple situation in which two frequencies $\omega$ and $\omega'$
are present on the two halves of the system, and take $\omega/\omega'$
to be a rational number, such that the Floquet theorem can be used.
As a first example, we take $\omega=2.7$ and $\omega'=\frac{\omega}{2}=1.35$
with time-independent parameters $A=-B=0.05,M=1,V_{0}=1$. In this
case $\omega$ is on resonance and $\omega'$ is nominally off-resonance.
However, $\omega'$ can induce two-photon resonances between valence
and conduction band states, and we find that in the $\omega'$ region
a gap opens as a result. Hence, the system is a Floquet insulator
throughout. Furthermore, we find that $C_{F}=1$ on both regions of
the sample. Yet, despite the fact that $C_{F}$ is constant, we obtain
a pair of modes with zero quasi-energy that are localized at the interface,
see Fig. \ref{Flo: HgTe Lin with space dependent w}. 

The mechanism behind this result is similar to that described Ref.
\cite{Tenebaum Katan  D. Podolsky} to explain the phase domain modes.
Since momentum $k_{x}$ in the direction parallel to the domain wall
is a good quantum number, the system can be analyzed for each $k_{x}$
value separately and one can define a $k_{x}-$dependent topological
invariant, $C_{k_{x}}^{'}$, as the winding number of $\hat{n}_{k}$
in the $(n_{k}^{y},n_{k}^{z})$ plane. We find by direct calculation
that $C_{k_{x}}^{'}$ has opposite signs in the $\omega$ and $\omega'$
regions. The localized states are therefore topologically protected,
provided particle-hole and reflection symmetries are present \cite{Tenebaum Katan  D. Podolsky}.

It is interesting to ask whether these effects can be seen in a scenario
where the frequency of the light varies continuously with position.
In practice, such a configuration could be realized, for example,
by passing a broad spectrum light source through a prism. To test
this, we chose a position-dependent frequency of the form $\omega(x)=\frac{\omega+\omega'}{2}+\frac{\omega-\omega'}{2}\tanh\left(\frac{x}{\lambda}\right)$,
such that the frequency interpolates between $\omega$ and $\omega'$
smoothly over a region of width $\lambda$. In this case, the Floquet
theorem no longer is valid, since the system is only quasi-periodic
and quasi-energies are no longer conserved. However, it is still possible
to evaluate the time evolution $U(T,0)$ for a very long time $T$
and search for localized eigenstates of $U$. In all cases that we
examined, we found that the localized interface modes survive when
$\lambda$ is smaller than the lattice constant, but that for larger
$\lambda$ these modes can no longer be discerned, indicating that
in practice this effect can only be seen for sharp jumps in frequency.
By contrast, by this procedure we find that the edge modes are robust
even when $\lambda$ is comparable with the system size. 

\subsection{Particle-hole symmetry breaking}

\begin{figure}[H]
\includegraphics[width=0.5\columnwidth]{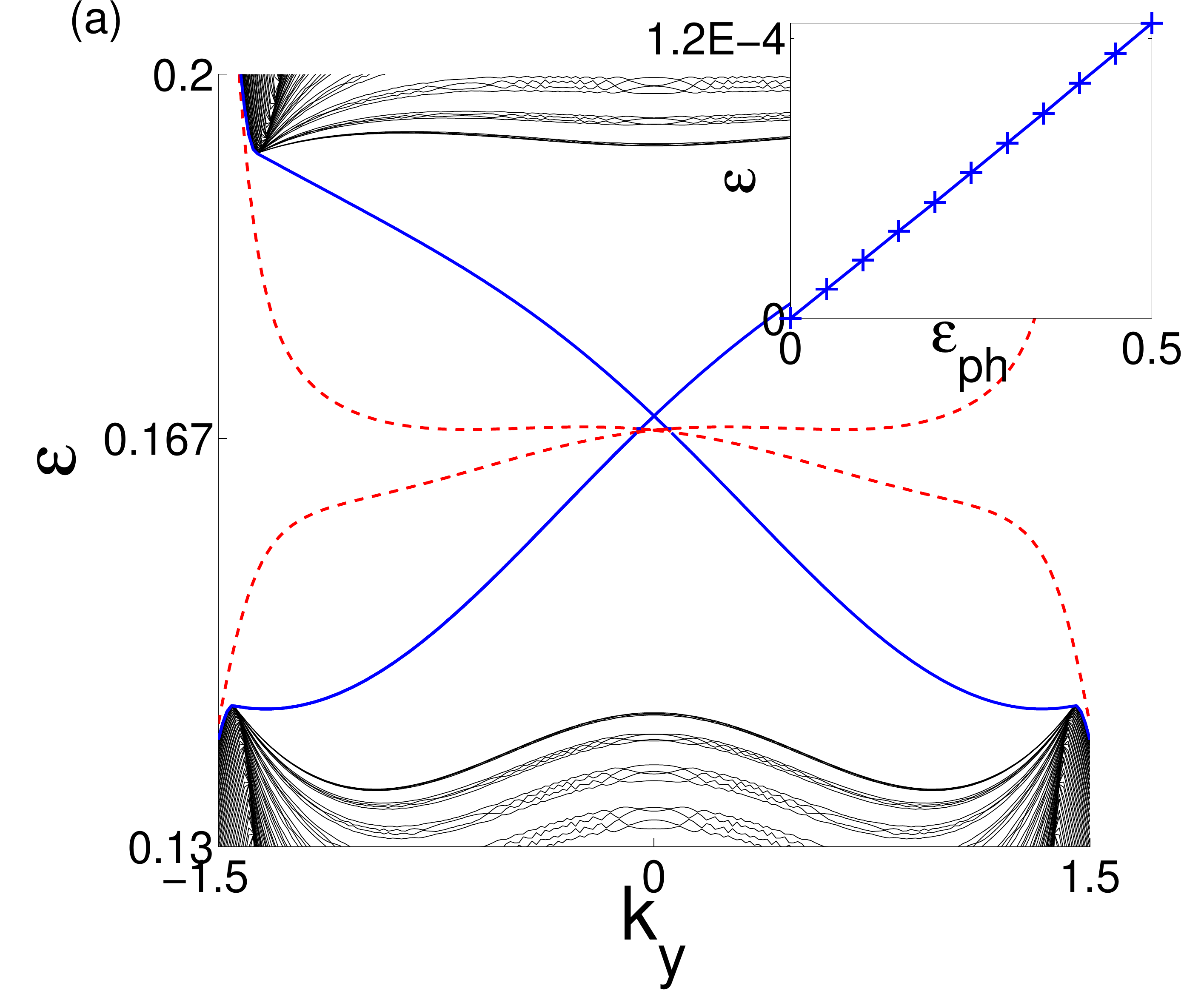}\includegraphics[width=0.5\columnwidth]{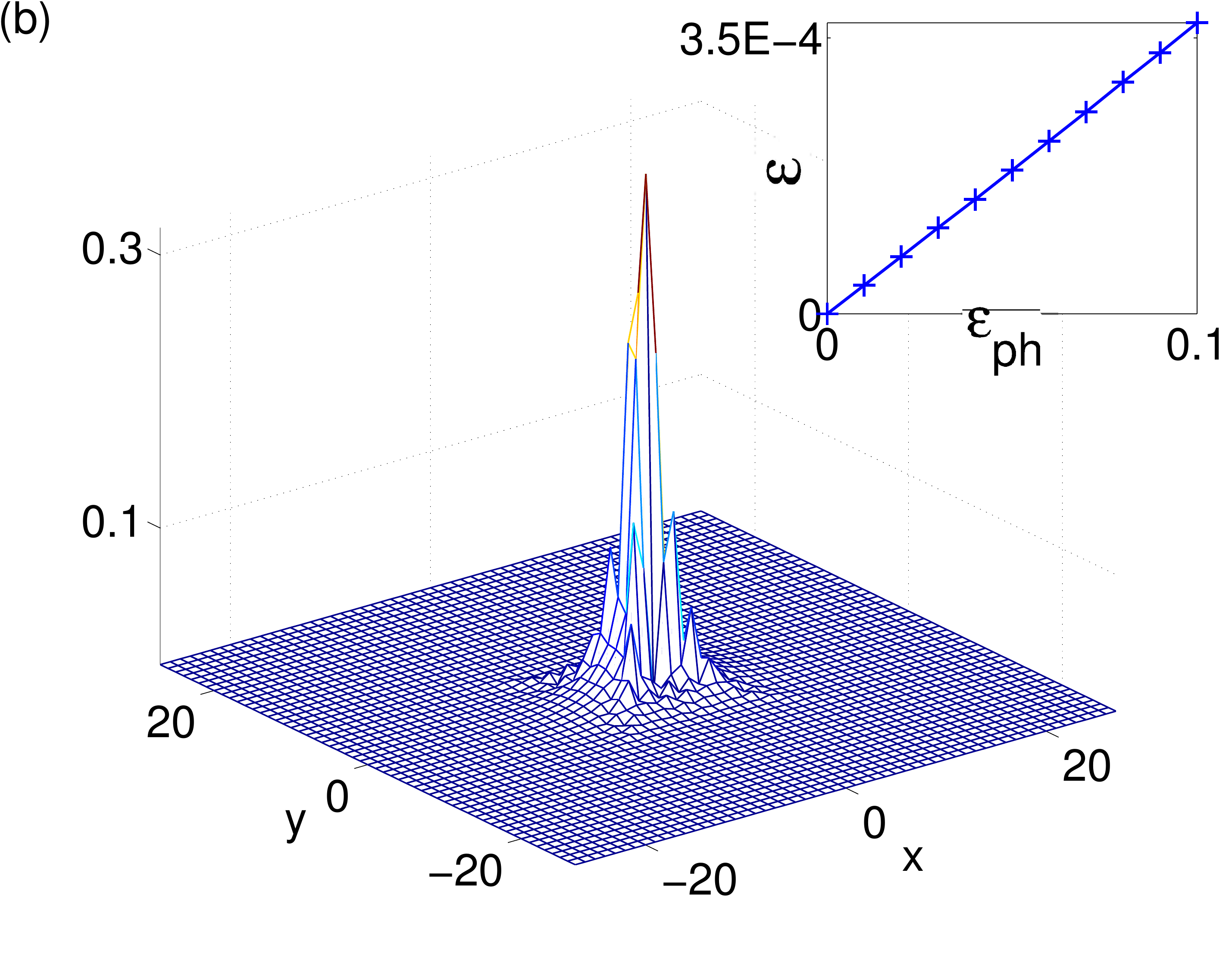}

\caption{The quasi-energy spectrum when PH and spatial inversion symmetries
are broken weakly\label{Flo:Numerical Spectrum SB}. Panel \textbf{(a)}
depicts the edge (solid blue) and domain wall (dashed red) modes for
$\epsilon_{ph}=0.1$. A small gap separates the domain wall states;
the inset depicts the dependence of the gap on $\epsilon_{ph}$. Panel
\textbf{(b)} displays the wave function amplitude of the vortex core
state for $\epsilon_{ph}=0.1$. The inset depicts the dependence of
the vortex core energy on $\epsilon_{ph}$. Results are for $A=-B=0.2,\: M=1,\:\omega=2.7$,
$V_{0}=0.4\:\left(1.5\right)$ and $L=200$ $\left(L=50\right)$ in
panel (a) (panel (b)). }
\end{figure}

The existence of quasi-stationary modes in the bulk of the zincblende
model was found to rely on the presence of PHS. However, in real systems
PHS is only approximate. Thus, it is natural to ask how these results
are affected by breaking of this symmetry. To answer this question,
we add a PHS breaking term
\[
\epsilon_{k}=-\epsilon_{ph}(\cos k_{x}+\cos k_{y}-2)
\]
and consider $\epsilon_{ph}\leq2|B|$, such that the time-independent
system remains gapped. Figure \ref{Flo:Numerical Spectrum SB}(a)
shows that localized sub-gap states survive the breaking of this symmetry,
and that the system remains topologically non-trivial. The edge modes
remain gapless in this case. However, a small gap opens in the Floquet
spectrum of the domain-wall modes. This gap is too small to be seen
in Fig. \ref{Flo:Numerical Spectrum SB}(a). The dependence of the
gap on $\epsilon_{ph}$ is shown in the inset of Fig. \ref{Flo:Numerical Spectrum SB}(a).
This small gap may be overridden by thermal fluctuations or a small
external bias. Thus, experiments carried out at temperatures above
this gap will not be sensitive to PHS breaking. 

The vortex core state in the vortex configuration shows higher degree
of robustness to PHS breaking than the domain wall modes. This robustness
is a result of the vortex core state being well separated in energy,
as well as in space, from the remaining of the spectrum. Specifically,
we find that weak PHS breaking shifts the quasi-energy of the bound
mode away from zero. However, the mode remains a mid-gap state that
is well separated from the remaining of the spectrum. Moreover, the
topological properties of this excitation, such as its charge, remain
fractional and are unaltered by breaking of PHS \cite{Tenebaum Katan  D. Podolsky}.

\section{Honeycomb Lattice Model \label{sub:Graphene}}

The generation of FTI is not limited to band-insulators. In particular,
previous work argued that on-resonance and off-resonance light can
both induce topological properties in graphene\cite{Fertig Gu Arovas Auerbach,Kitagawa_Oka_Brataas_Fu_Demler}
and in photonic crystals based on honeycomb lattices \cite{Photonic Crystals 2}.
From an experimental point of view, Floquet topological insulators
have been realized in photonic crystals for a honeycomb lattice with
helical wave guides, see Ref. \cite{Photonic Crystal}.

We will now demonstrate that the modulated FTI can be realized in
a honeycomb lattice. We model the system by the tight binding approximation
\cite{Castro Neto}. The Hamiltonian is given by Eq. (\ref{eq:H2x2})
with $\varepsilon_{k}=0$ and $\vec{d}_{k}=t\left(-\Re f_{k},\Im f_{k},0\right)$,
where $f_{k}=1+2e^{i\frac{3}{2}k_{y}a}\cos\left(\frac{\sqrt{3}}{2}k_{x}a\right)$.
Here, $a$ is the inter-atomic distance and $t$ is the hopping parameter.
The spectrum consists of conduction and valence bands with Dirac cones
that intersect at two Fermi points, $K_{\pm}=\left(\frac{2\pi}{3\sqrt{3}a},\pm\frac{2\pi}{3a}\right)$.
$K_{\pm}$ are commonly referred to as ``valleys'', and are related
to each other by time reversal. Note that $\left|f_{k}\right|^{2}\leq3t$,
such that the bandwidth is $6t$. The system is invariant under particle-hole,
spatial inversion, and time reversal symmetries \cite{Castro Neto}. 

The generation of topological insulators requires the formation of
a gap in the energy spectrum. In the studied model, the two Fermi
points are protected by spatial inversion and time-reversal symmetries.
The formation of FTI therefore requires the breaking of either of
these symmetries. As a result, linearly polarized light which does
not break TR nor spatial inversion cannot induce a gap or generate
topological properties. By contrast, circularly polarized light induces
a gap of opposite masses on each of the two Dirac cones, thus creating
a topological Floquet spectrum \cite{Torres,Fertig Gu Arovas Auerbach,Kitagawa_Oka_Brataas_Fu_Demler,Oka_Aoki,Torres2,Roslyak_Gumb_Huang}.

\subsubsection{Off-Resonance Light\label{sub:Off-Resonant-light}}

\begin{figure}
\includegraphics[width=0.7\columnwidth]{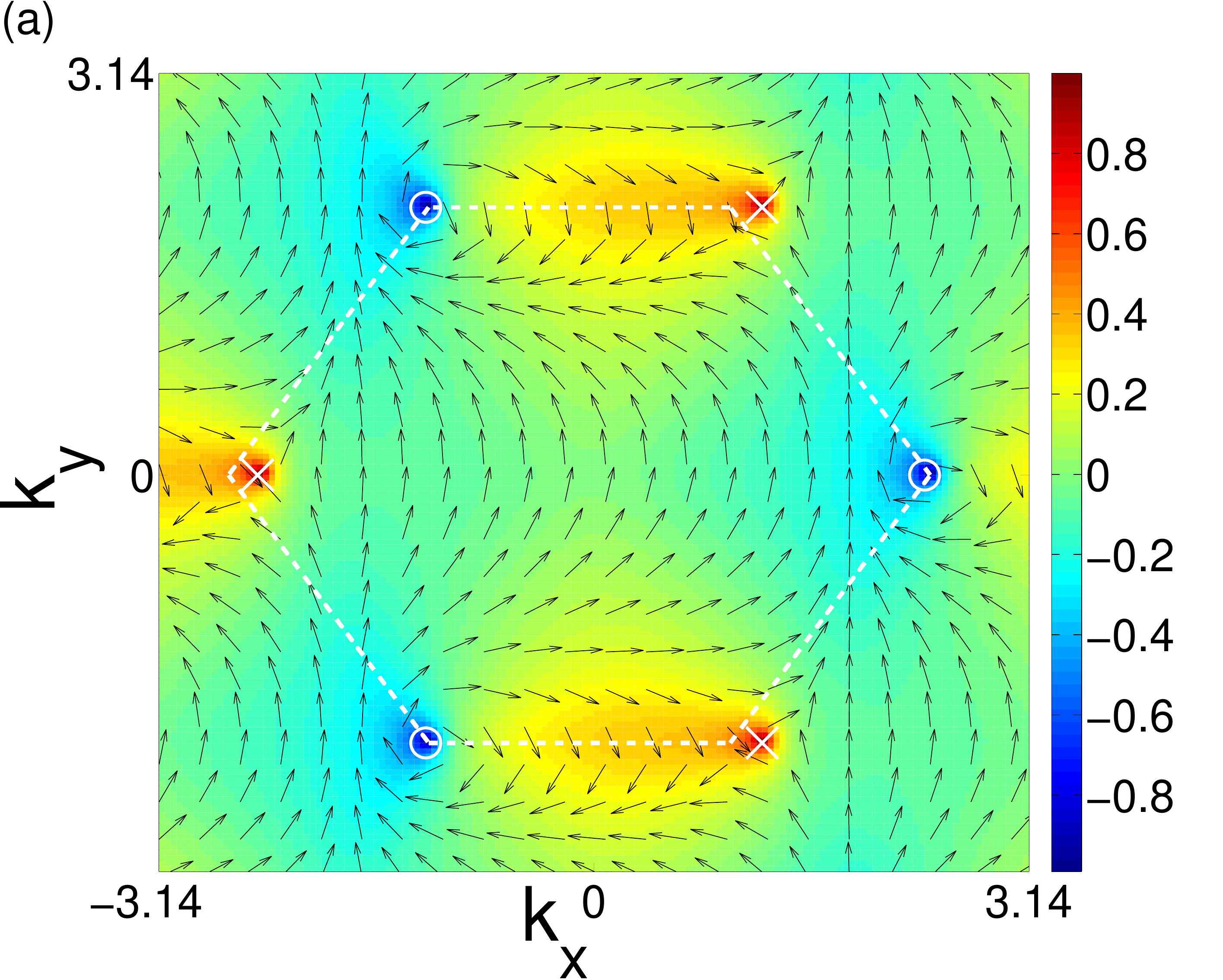}

\includegraphics[width=0.7\columnwidth]{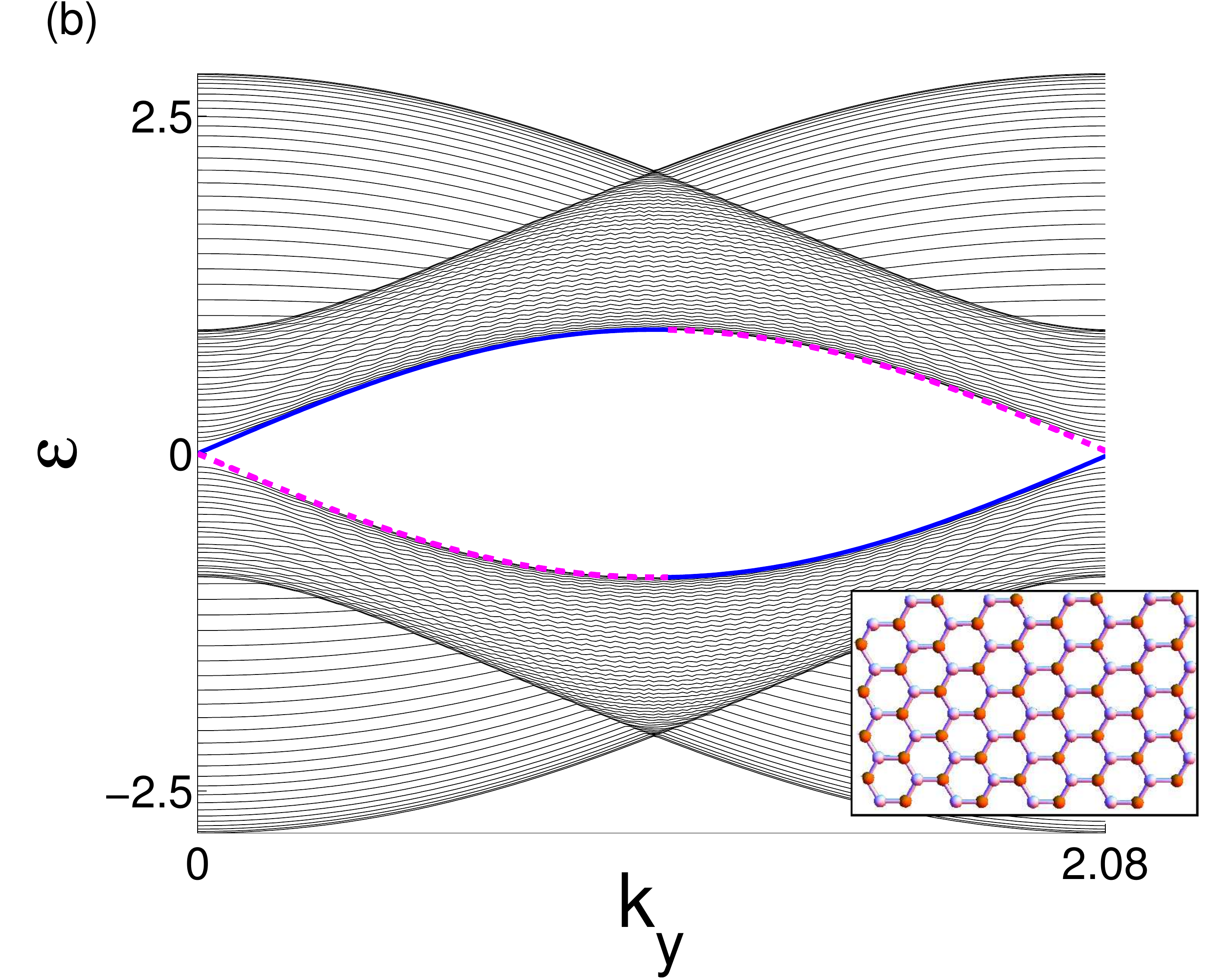}

\caption{\textbf{(a)} $\hat{n}_{k}$ over the first Brillouin zone for the
honeycomb lattice model irradiated \label{Flo:n(k) graphene} by left
handed polarized light with off-resonance frequency $\left(\omega=6.5t\right)$.
The colors denote the magnitude of $\hat{n}_{z}$ and the arrows are
the direction of $\hat{n}_{x}$ and $\hat{n}_{y}$. The $X$ markers
denote north poles and the circles denote south poles. There is one
north pole, with vorticity 1 and one south pole with vorticity -1.
This corresponds to $C_{F}=1$. \textbf{(b)} The Floquet spectrum.
The blue (dashed green) lines correspond to states localized at the
right (left) edge.\textbf{ }The inset gives an illustration of a honeycomb
lattice in an Armchair configuration}
\end{figure}

\begin{figure}
\includegraphics[width=0.5\columnwidth]{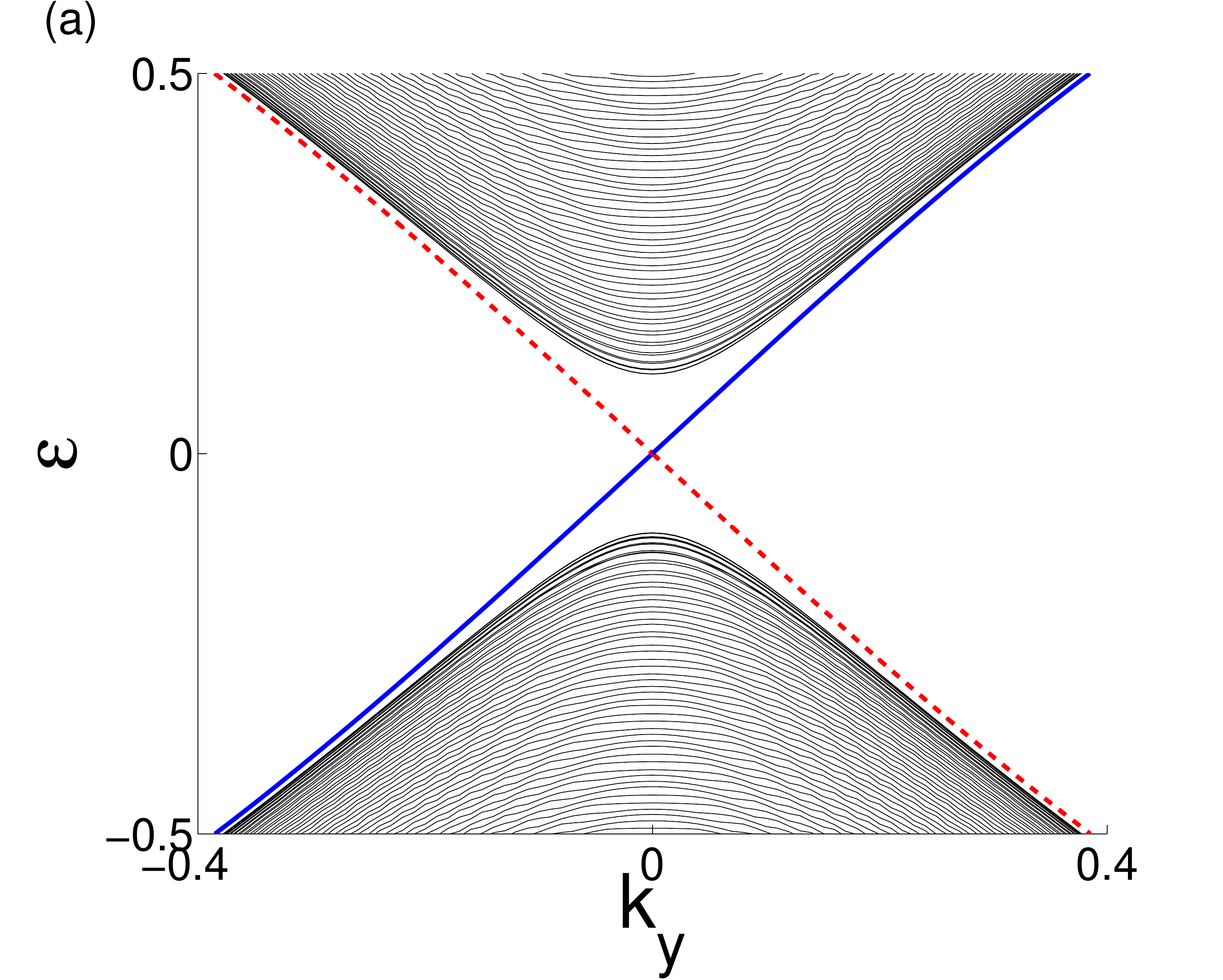}\includegraphics[width=0.5\columnwidth]{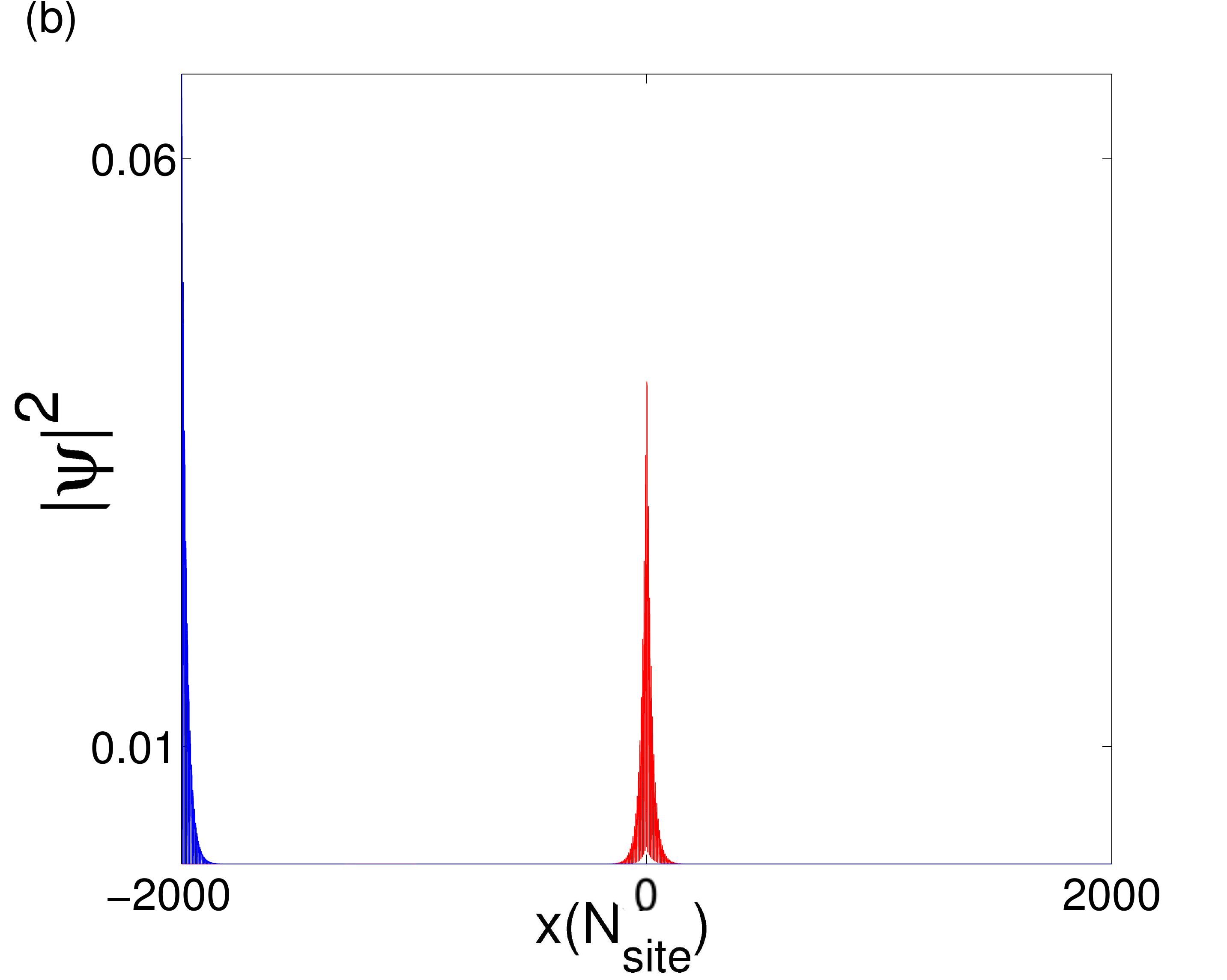}\caption{\textbf{(a)} Floquet spectrum of the honeycomb lattice with off-resonance
light for a RH/LH interface\label{Flo: Grpahene Circ. Off-Res. Domain Wall}.
The dispersion of the two edge modes (solid blue) overlaps, and so
does the dispersion of the two bulk modes (dashed red). Results are
for $\omega=6.5,A_{0}=0.5,L=200$ \textbf{(b)} Amplitude of one of
the domain wall modes (red) and one of the edge modes (blue) for $\omega=6.5,A_{0}=0.5,L=2000.$}
\end{figure}

Let us first examine the effect of uniform light with off-resonance
frequency $\left(\omega>6t\right)$ \cite{Kitagawa_Oka_Brataas_Fu_Demler}.
We solve for the Floquet Hamiltonian with periodic boundary conditions.
Figure \ref{Flo:n(k) graphene} shows that for left circularly polarized
light, $\hat{n}_{k}$ points towards the north pole at the $K_{-}$valley
and towards the south pole at the $K_{+}$ valley. Surrounding these
points, the $x$ and $y$ components of $\hat{n}_{k}$ form a vortex
and an anti-vortex, respectively. This indicates that the Dirac points
pick up opposite masses, and that the Chern number is $C_{F}=1$.
Similarly, for right-polarized light $C_{F}=-1$, as expected from
TRS invariance.

The topological nature of the system is reflected in the presence
of chiral edge modes, as illustrated in Figure \ref{Flo:n(k) graphene},
which shows that the spectrum for a ribbon in the armchair configuration.
The spectrum includes two quasi-stationary modes, localized at each
of the two edges. Similar results are obtained for a ribbon in a zig-zag
configuration.

Motivated by earlier results, we searched for quasi-stationary states
both at a domain wall and also at a vortex in the \emph{phase} of
the light. However, we found through numerical simulations that when
the light is off-resonance, neither one of these configurations induce
quasi-stationary states in the bulk.

In order to explain these results, we evaluate the Floquet Hamiltonian
analytically near the Dirac points. The low energy Hamiltonian of
the honeycomb lattice model consists of two blocks, each describing
the system near a valley ($\tau_{z}=\pm1),$

\begin{equation}
H_{k}=-v_{f}\left(\sigma_{x}k_{x}+\sigma_{y}\tau_{z}k_{y}\right)\label{eq:Graphene H0 lowE}
\end{equation}
where $v_{f}=\frac{3}{2}ta$ is the Fermi velocity. We describe the
effect of the radiating light by the minimal substitution, $H_{\vec{k}}\to H_{\vec{k}+\vec{A}}$,
where $\vec{A}\left(t\right)=A_{0}\left(\cos\left(\omega t+\alpha\right),\sin\left(\omega t+\alpha\right),0\right)$
for LH polarized light.

The effect of off-resonance light (see \cite{Kitagawa_Oka_Brataas_Fu_Demler})
can be described by a static Floquet Hamiltonian, with a Floquet spectrum
described in terms of a ``dressed'' energy spectrum. In this case,
the RWA can not be used to estimate the Floquet Hamiltonian and a
different approach is required. We use perturbation theory and expand
the time evolution operator, Eq. (\ref{eq:Udef}), as a series in
the small parameter $\frac{A_{0}^{2}}{\omega}$. For uniform light,
this procedure yields

\begin{equation}
\begin{array}{c}
H_{F}=H_{0}+\frac{1}{\omega}\left[H_{1},H_{-1}\right]\\
+\frac{1}{\omega}\left(e^{i\alpha}\left[H_{0},H_{1}\right]-e^{-i\alpha}\left[H_{0},H_{-1}\right]\right)+\mathcal{O}\left(\frac{A_{0}^{3}}{\omega^{2}}\right)
\end{array}\label{eq:Hf perturbation theory}
\end{equation}
where $H_{n}=\frac{1}{\tau}\int_{0}^{\tau}e^{-i\omega nt}H\left(t\right)dt$
is the $n^{th}$ Fourier coefficient of the Hamiltonian. Thus $H_{0}$
is given by Eq. (\ref{eq:Graphene H0 lowE}) and
\[
H_{\pm1}=A_{0}\left(\sigma_{x}\pm i\sigma_{y}\right)
\]
6aThus, the Floquet Hamiltonian is
\begin{equation}
\begin{array}{c}
H_{F}=-v_{f}\left(\sigma_{x}k_{x}+\tau_{z}\sigma_{y}k_{y}\right)+\tau_{z}\sigma_{z}m\\
\qquad-\tau_{z}\sigma_{z}\Sigma_{k,\alpha}+\mathcal{O}\left(\frac{A_{0}^{3}}{\omega^{2}}\right)
\end{array}\label{eq:Hf off-resonance}
\end{equation}
where $m=v_{f}^{2}\frac{A_{0}^{2}}{\omega}$ is the effective mass
at the two valleys and $\Sigma_{k,\alpha}=v_{f}A_{0}\left(k_{x}\cos\alpha+k_{y}\sin\alpha\right)$
is a linear term which can be absorbed into an $\alpha$-dependent
shift in the position of the Dirac points. Note that the Floquet Hamiltonian
is otherwise independent of $\alpha$. As a result, no localized modes
are generated by space modulation of the phase $\alpha$. In a similar
manner, since modifying $\omega$ does not affect the sign of $m$
provided $\omega$ is off-resonance, no quasi-stationary modes are
induced by light radiation with space-modulated frequency. 

By contrast, if we make a domain wall in the \emph{polarization} of
the light, that is, an interface between left and right polarizations,
this results in two chiral modes at the interface that propagate in
the same direction. This can simply be understood in terms of the
change in Chern number between the two domains, as shown in Fig. \ref{Flo: Grpahene Circ. Off-Res. Domain Wall}.
Such a configuration can be created, for instance, in a photonic crystal,
by using wave guides with opposite helicities. 

In Ref. \cite{Tenebaum Katan  D. Podolsky} we found that imposing
a slowly-varying phase twist results in a current analogous to the
Josephson effect, $j=\rho_{s}\nabla\alpha$ . This result also holds
here, where we find numerically that a slow modulation in space of
$\alpha$ yields a DC current proportional to $\nabla\alpha$.

\subsubsection{On-Resonance Light}

We will now extend the analysis to on-resonance light. For concreteness
we will consider light frequencies in the range $3t<\omega<6t,$ that
is, frequencies that only allow a single photon resonance. The frequencies
of the radiating light in this scenario are small compared to the
off-resonance case. This enables the use of less energetic photons
in experiments, thus opening a possibility for additional realizations
of our results in condensed matter systems. For example, recent work\cite{Torres}
demonstrated that resonant light can induce controlled gapless modes
in graphene.

The Chern number for left-polarized light is $C_{F}=3$, as seen from
the winding number of $\hat{n}_{k}$ in Figure \ref{Flo: Grpahene on-resonance Spectrum, Uniform Light}.
The change in Chern number relative to the off-resonance case is due
to the folding in energy for states with momenta on a circle around
the $\Gamma=(0,0)$ point. We demonstrate this topological nature
by evaluating the spectrum on a ribbon in an armchair configuration.
Figure \ref{Flo: Grpahene on-resonance Spectrum, Uniform Light} shows
that three quasi-stationary modes are now localized at each boundary.
Focusing on one edge, there are two right-moving modes with quasi-energy
$\frac{\omega}{2}$ and one left-moving mode with zero quasi energy. 

\begin{figure}
\includegraphics[width=0.7\columnwidth]{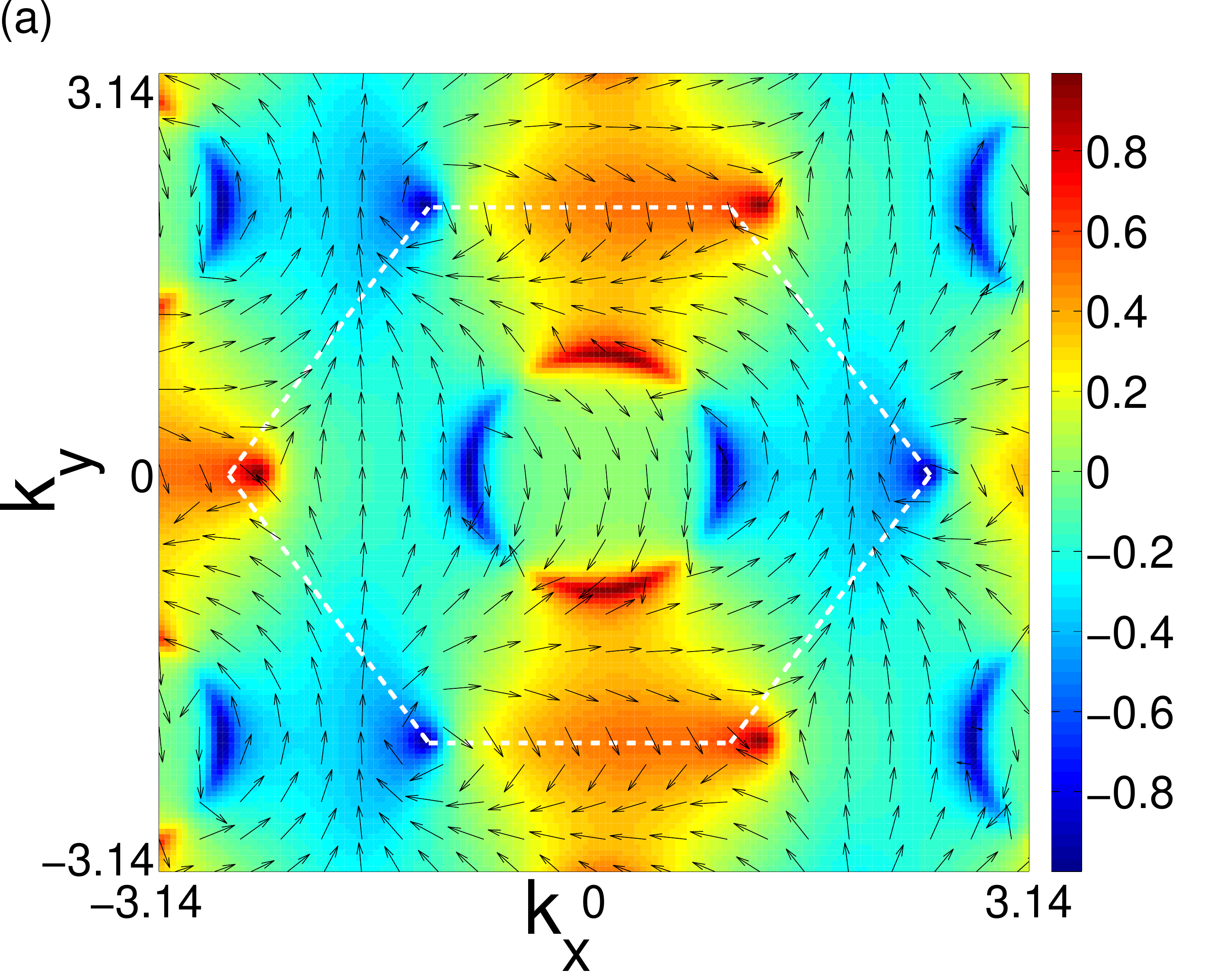}

\includegraphics[width=0.7\columnwidth]{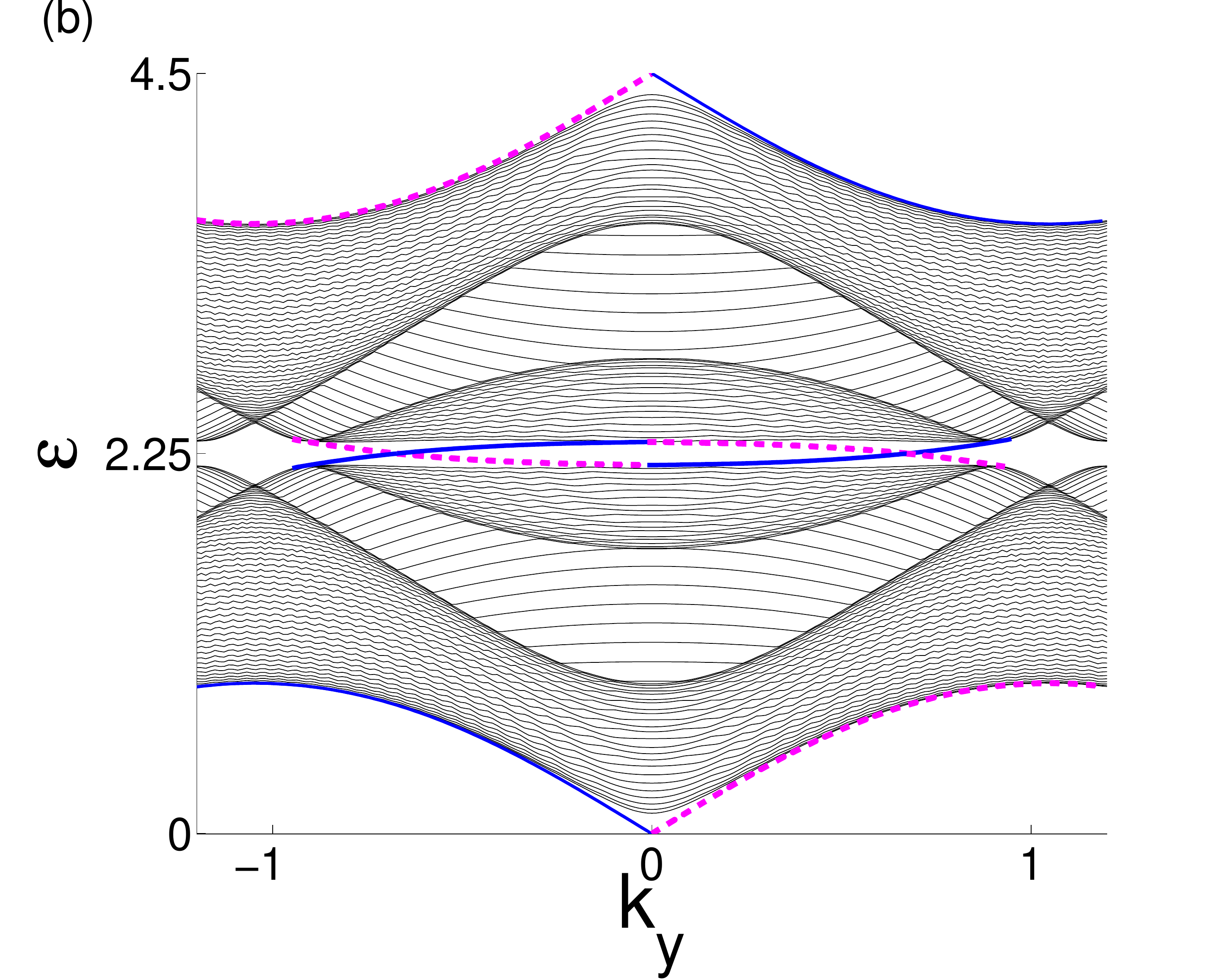}\caption{\label{Flo: Grpahene on-resonance Spectrum, Uniform Light}\textbf{
}The spectrum of the honeycomb lattice model irradiated by resonant
light. Results are for $\omega=4.5\:,A_{0}=0.5,\: L=200$\textbf{
(a)} $\hat{n}_{k}$\textbf{ }over the first Brillouin zone. The color
map shows the magnitude of $\hat{n}_{z}$, and the arrows are the
direction of $\hat{n}_{x}$ and $\hat{n}_{y}$. The dashed line denotes
the first BZ, and the $X$ and $O$ markers denote north and south
poles, respectively. There are in total three north poles, all with
positive vorticity, corresponding to $C_{F}=3$. \textbf{(b) }The
Floquet spectrum. The blue (dashed brown) lines correspond to states
localized at the right (left) edge.}
\end{figure}

\begin{figure}
\includegraphics[width=0.5\columnwidth]{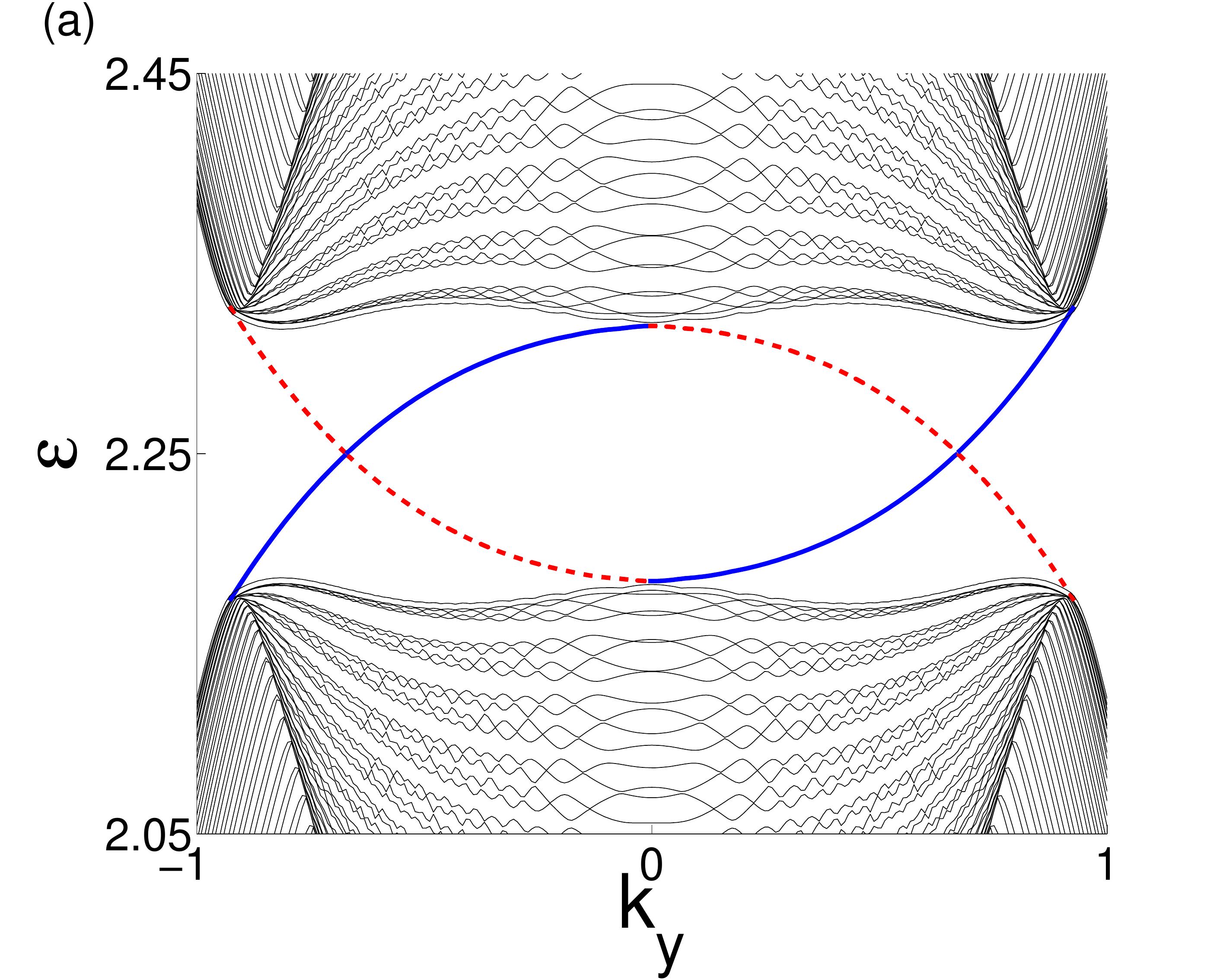}\includegraphics[width=0.5\columnwidth]{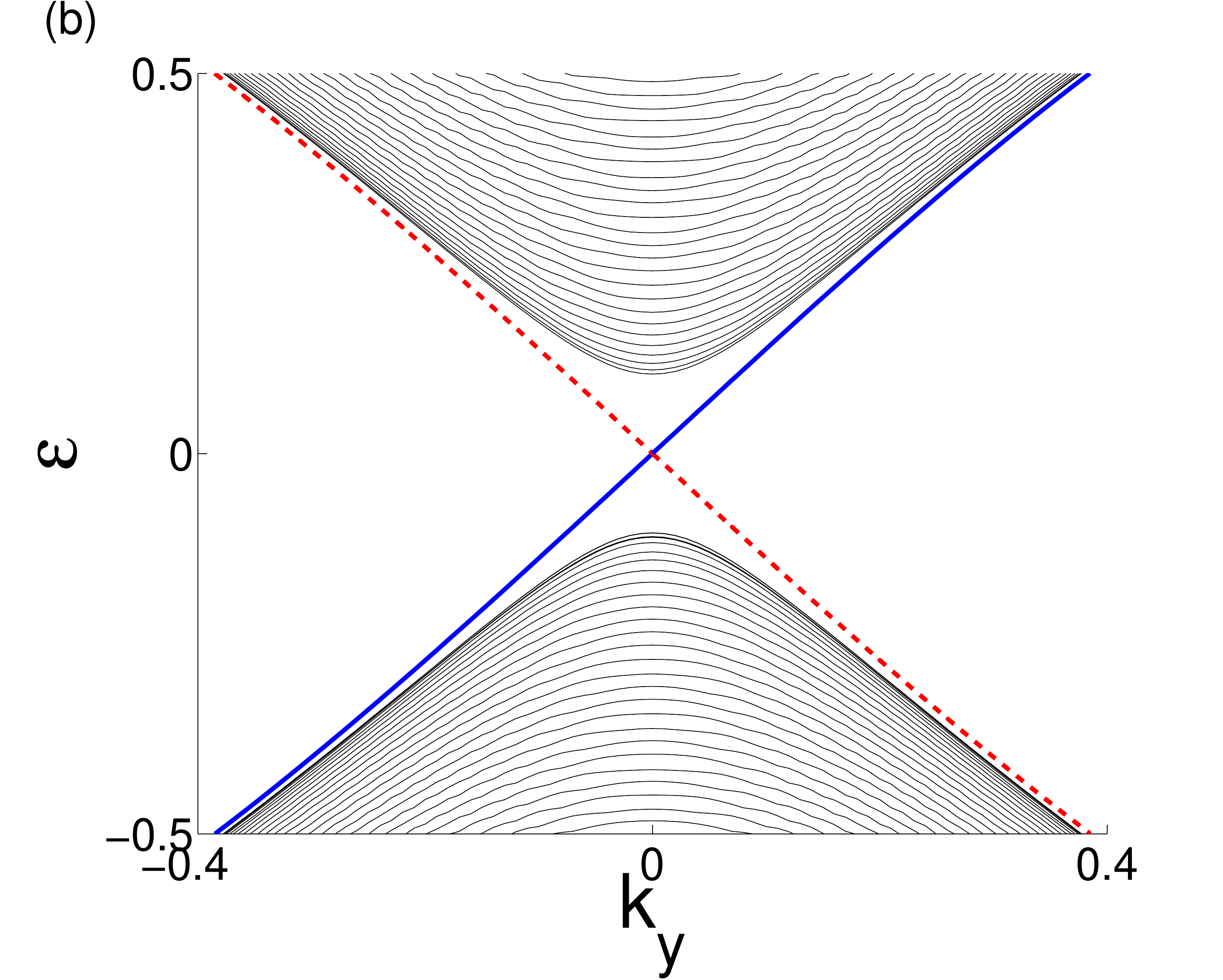}\caption{\label{Flo: Grpahene Circ. On-Res. Domain Wall and LH/RH} Floquet
spectrum of irradiated honeycomb lattice, for resonant light with
space-modulated polarization. A domain wall separates left and right
polarized light. Blue lines denote edge modes and dashed red lines
denote bulk modes \textbf{(a) }Two doubly degenerate bulk modes and
two doubly degenerate edge modes have $\varepsilon=\frac{\omega}{2}$.
\textbf{(b)} Two localized modes have zero quasi-energy. These states
have identical low energy group velocity. In total there are 6 localized
modes, in correspondence to $\Delta C_{F}=6$ across this interface.
Results are for $A_{0}=0.5t,\,\omega=4.5t,\, L=500$}
\end{figure}

\begin{figure}
\includegraphics[width=0.5\columnwidth]{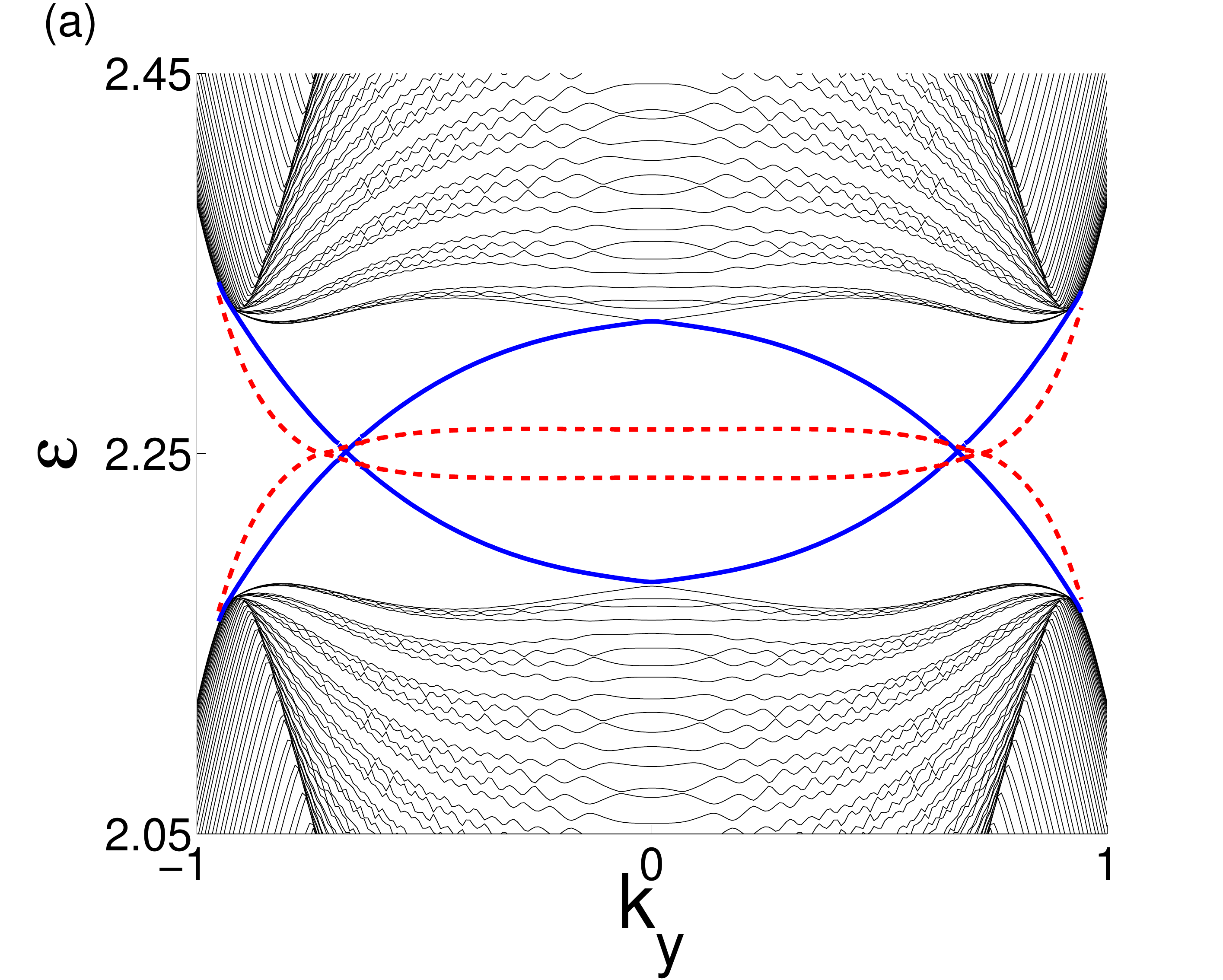}\includegraphics[width=0.5\columnwidth]{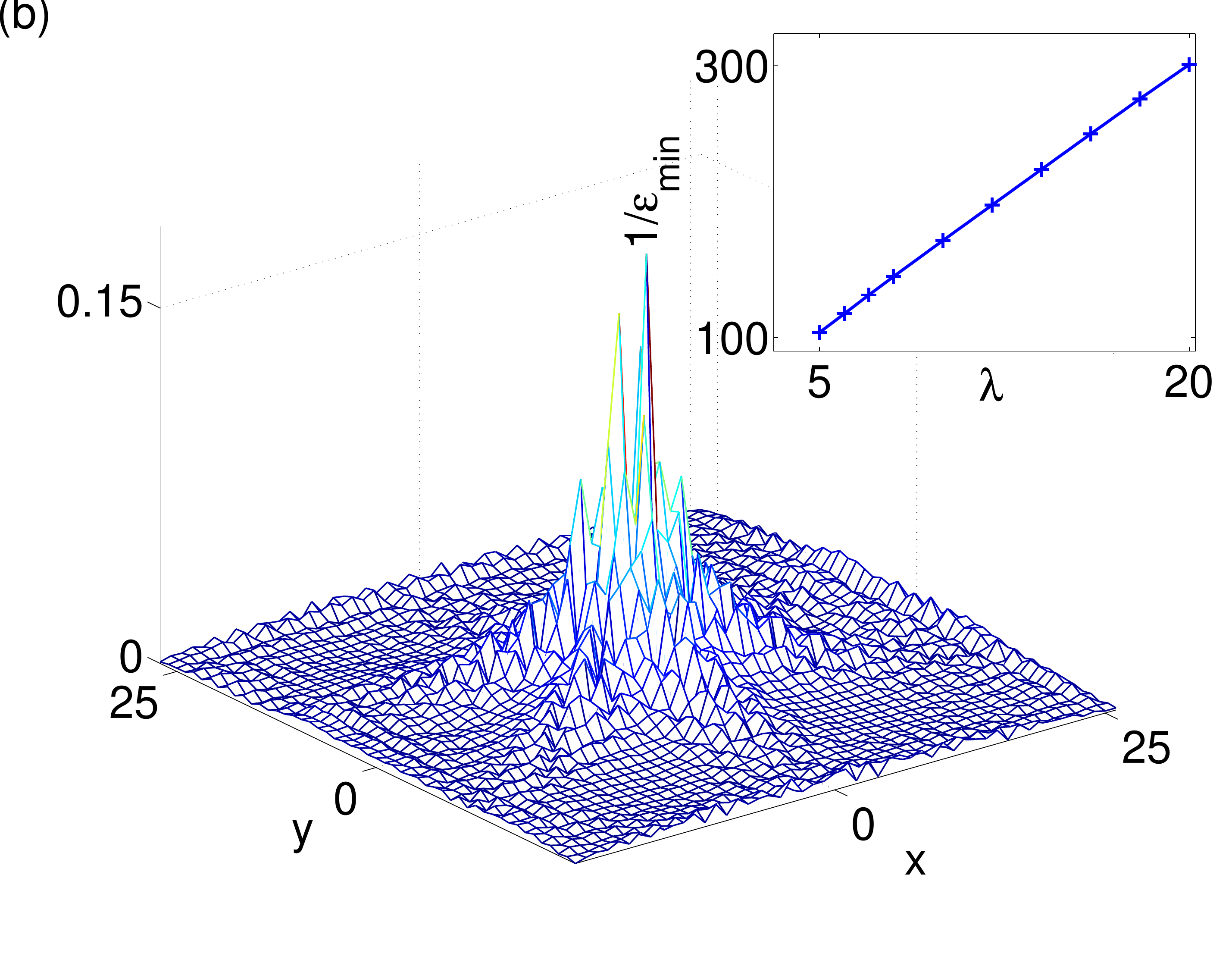}\caption{\label{Flo: Grpahene Circ. On-Res, vortex}\textbf{(a) }The dispersion
of the domain wall modes (dashed red line) and edge states (solid
blue line).\textbf{ }While the edge modes are gapless, the bulk modes
have an energy gap which decreases exponentially with $\lambda$.
Results are for $A_{0}=0.5t,\omega=4.5t$\textbf{ (b)} The wave function
amplitude of a vortex core state for a smooth vortex with $\lambda=18$.
The inset shows the energy of the lowest lying vortex-core state as
a function of $\lambda$. Results are for $A_{0}=0.5t,\omega=3t$}
\end{figure}

We now allow the light to vary in space. As a first example, we examine
an interface between right and left handed polarized light. As in
the off-resonance case, the spectrum of this setup includes gapless
domain modes, which are a direct result of the change in Chern numbers
at the interface. In this case, we find three right movers and three
left movers at the interface, see Fig. \ref{Flo: Grpahene Circ. On-Res. Domain Wall and LH/RH}.

Next, we consider a vortex configuration, with $\alpha=\arctan\left(\frac{y}{x}\right)$.
In this configuration, the phase $\alpha$ winds by $2\pi$ about
the origin, which is taken to be at the center of the sample. Numerical
simulations show that in addition to the edge modes, the Floquet spectrum
now includes pairs of positive and negative quasi-energy states, localized
at the vortex core. These modes have finite energy\textbf{.} For example,
for $A_{0}/t=0.5$ and $\omega/t=3$, we find \textbf{$E_{vortex}=3\times10^{-2}$}.
Furthermore, we find that for a smoothly varying vortex configuration,

\begin{equation}
\vec{A}=A_{0}\tanh\left(\frac{r}{\lambda}\right)\left(\cos\left(\omega t+\alpha\right),\sin\left(\omega t+\alpha\right),0\right),\label{eq:A0,tanh}
\end{equation}
where $\alpha=\arctan\left(y/x\right),\: r=\sqrt{x^{2}+y^{2}}$ and
$\lambda$ is the the size of the vortex core, the energy of the vortex
core modes decreases as $1/\lambda$. For example, for $\lambda=14a$
and all other parameters as above, we find\textbf{ $E_{vortex}=4\times10^{-3}$.
}Note that in a physical setup, $\lambda$ will be of order of the
wavelength of the light, which is much larger than the values of $\lambda$
accessible in our simulations. The value of the energy in a physical
setup will be correspondingly smaller.

As a third example, we consider a domain wall configuration in which
$\vec{A}\left(t\right)$ flips sign at $y=0$. This corresponds to
a $\pi$ phase shift in the light, without a change in the polarization.
As in the vortex configuration, the spectrum includes localized states
with finite energy gap. If instead the domain wall is smoothened,
$\vec{A}=A_{0}\tanh\left(\frac{x}{\lambda}\right)(\cos\omega t,\sin\omega t,0),$
the energy gap decays exponentially with $\lambda$, as shown in Fig.
\ref{Flo: Grpahene Circ. On-Res, vortex}. For example, for $\lambda=14a$
we find $E_{domain-wall}=6\times10^{-10}$.

In order to explain these results we evaluate the Floquet Hamiltonian
around the $\Gamma$ point. For resonant light, we can use the RWA
for this purpose. We find that up to a unitary transformation ($\psi\rightarrow e^{ik_{y}\sigma_{z}}\psi$)
the Floquet Hamiltonian is (in the $\left(\hat{z},\hat{x},\hat{y}\right)$
coordinate basis) 
\begin{equation}
\begin{array}{c}
H_{F}^{\Gamma}=3t\left(\begin{array}{cc}
\frac{k^{2}}{2m}-\mu & \Delta_{k}e^{-i\alpha}\\
\Delta_{k}e^{i\alpha} & -\frac{k^{2}}{2m}+\mu
\end{array}\right)\end{array}\label{eq:Graphene Hf}
\end{equation}
where $\Delta_{k}=\frac{A_{0}}{16}\left(k_{x}-ik_{y}\right)^{2}$,
$m^{-1}=\frac{1}{4}\left(\frac{\omega}{t}-3\right)$ and $\mu=1-\frac{\omega}{6t}$.
Equation (\ref{eq:Graphene Hf}) is analogous to a $d+id$ SC. For
$\omega<6t$, the chemical potential is positive and $H_{F}^{\Gamma}$
is topological. It supports two topologically protected chiral modes
at the system edges. These modes have quasi-energy $\omega/2$, as
opposed to the mode arising due to the gap at the valleys, which has
quasi-energy $0$. When $\omega>6t$ , the chemical potential becomes
negative and Eq. (\ref{eq:Graphene Hf}) corresponds to a trivial
insulator. Then, only the mode due to the gap at the valleys survive.

The existence of the quasi-stationary modes can be understood through
the $d+id$ SC analogy. In the vortex configuration, a vortex core
state is analogous to the $d+id$ SC Caroli-de-Gennes-Matricon vortex
states (see Ref. \cite{Caroli}), with an energy gap that is proportional
to $\frac{1}{\lambda}$.

In the domain wall configuration, the BdG equations have four quasi-stationary
solutions. These modes are analogous to the d-wave $\pi$-junction
modes, see Ref. \cite{Josephson in P-wave}, with linear low energy
dispersion. We find that in the step function configuration these
states have a finite gap, while in the smooth domain wall setup the
gap decays exponentially with $\lambda$.

\section{Three dimensional systems\label{sub:3D-Topological-Insulator}}

In this section we generalize our results to a three dimensional model.
In the case of 3d FTI, the localized surface modes form an odd number
of Dirac cones. Following Ref. \cite{Lindner3d}, we first present
a brief description of the formalism for a 3D FTI. Consider a general
$4\times4$ trivial insulator, described by the Hamiltonian 
\begin{equation}
\begin{array}{c}
H_{k}=D_{k}^{\mu}\gamma_{\mu}+\varepsilon_{k}I\end{array}.\label{eq:3D H unperturbed}
\end{equation}
Here, $D_{k}^{\mu}$ is a 4-vector that depends on the 3d lattice
momentum $k$, and $\gamma^{\mu}=\left(\gamma^{1},\gamma^{2},\gamma^{3},\gamma^{4}\right)$
is a vector composed of the four Dirac matrices, given by $\gamma_{i}=\sigma_{i}\otimes\tau_{x}$
for $i=1,2,3$ and $\gamma_{4}=I\otimes\tau_{z}$. In addition, we
define $\gamma_{5}=I\otimes\tau_{y}$. This Hamiltonian describes
an insulator provided that $D_{k}^{\mu}$ does not vanish anywhere
on the Brillouin zone. We take $D_{k}^{\mu}=(\vec{d}_{k},D_{k}^{4})$
with $\vec{d}_{k}$ odd and $D_{k}^{4}$ even under spatial inversion.
Then, Eq. (\ref{eq:3D H unperturbed}) is invariant under both time
reversal and space inversion symmetries. For simplicity we first consider
the situation where there is particle-hole symmetry, such that $\varepsilon_{k}=0$. 

According to the topological classification of 3d systems \cite{Schneider},
time-reversal topological insulators are characterized by a $\mathbb{Z}_{2}$
topological invariant. In the case of Eq. (\ref{eq:3D H unperturbed}),
a natural topological invariant is given by a generalization of the
TKNN formula to higher dimensions. Here, the Brillouin zone is a three
dimensional torus, $T{}^{3}$, and $\hat{D}_{k}$ is a four component
unit vector that lies on $S{}^{3}$. $\hat{D}_{k}$ can therefore
be described as a map, $D:T^{3}\rightarrow S^{3}$ . We define the
topological invariant $\chi$ as the degree of this map, which is
the number of times $\hat{D}_{k}$ wraps around the unit sphere $S^{3}$
as $\vec{k}$ runs over the Brillouin zone. When PH and TR symmetries
are both present $\chi$ is integer valued. When PH breaking terms
are included, $\chi$ is only defined modulo 2.

As an example of a 3d FTI, we consider a cubic lattice model, with
\[
\begin{array}{c}
\vec{d}_{k}=A\left(\sin k_{x},\sin k_{y},\sin k_{z}\right)\\
D_{k}^{4}=M+2B\left(\cos k_{x}+\cos k_{y}+\cos k_{z}-3\right).
\end{array}
\]
Similarly to its two dimensional counterpart, discussed in Sec. \ref{sec:Square-lattice-model},
the topology of the unperturbed system depends on the choice of parameters.
We take $M/B<0$, in which case the time-independent system is a trivial
insulator with a doubly degenerate spectrum.

\begin{figure}[t]
\includegraphics[width=0.485\columnwidth]{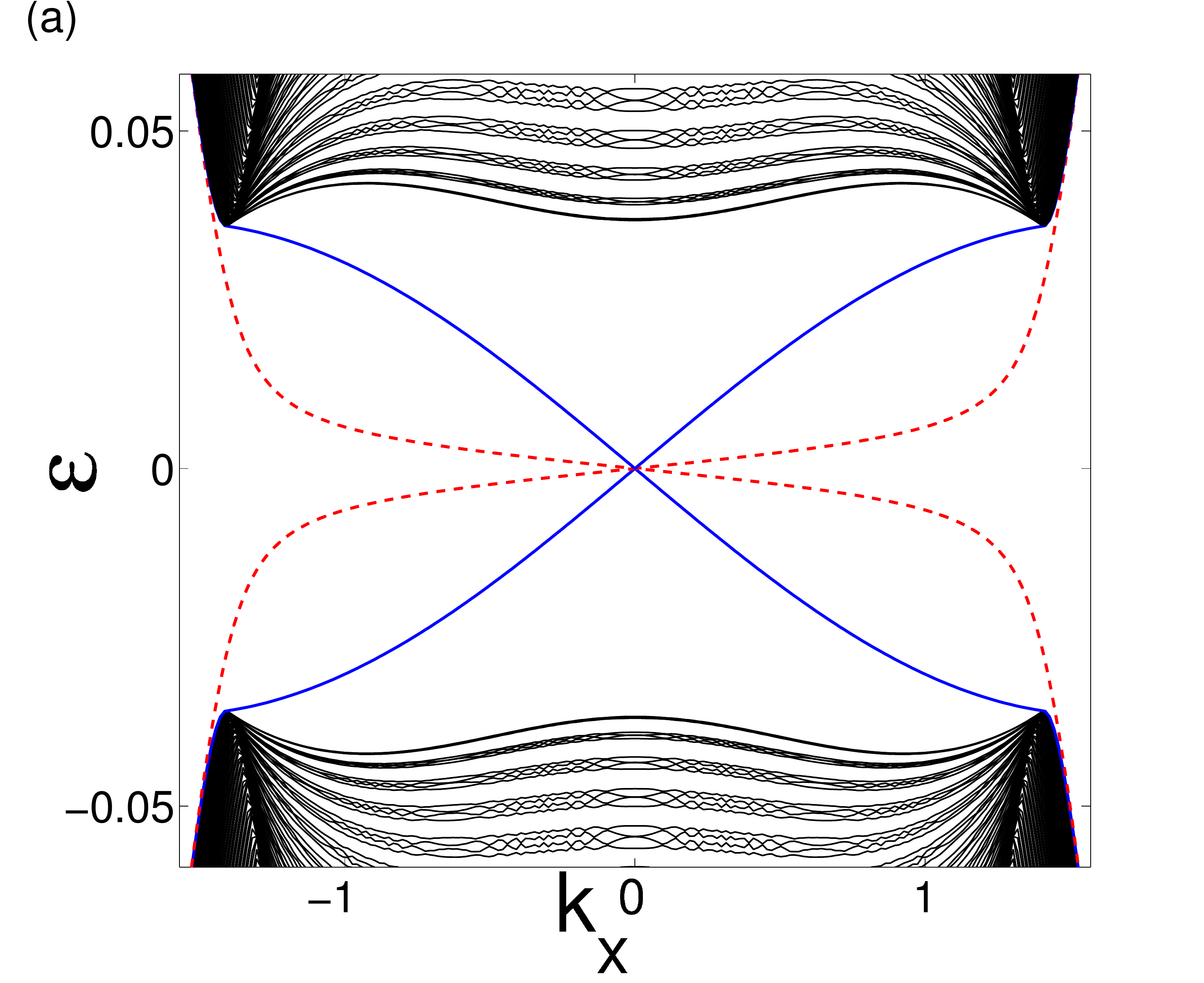}\includegraphics[width=0.515\columnwidth]{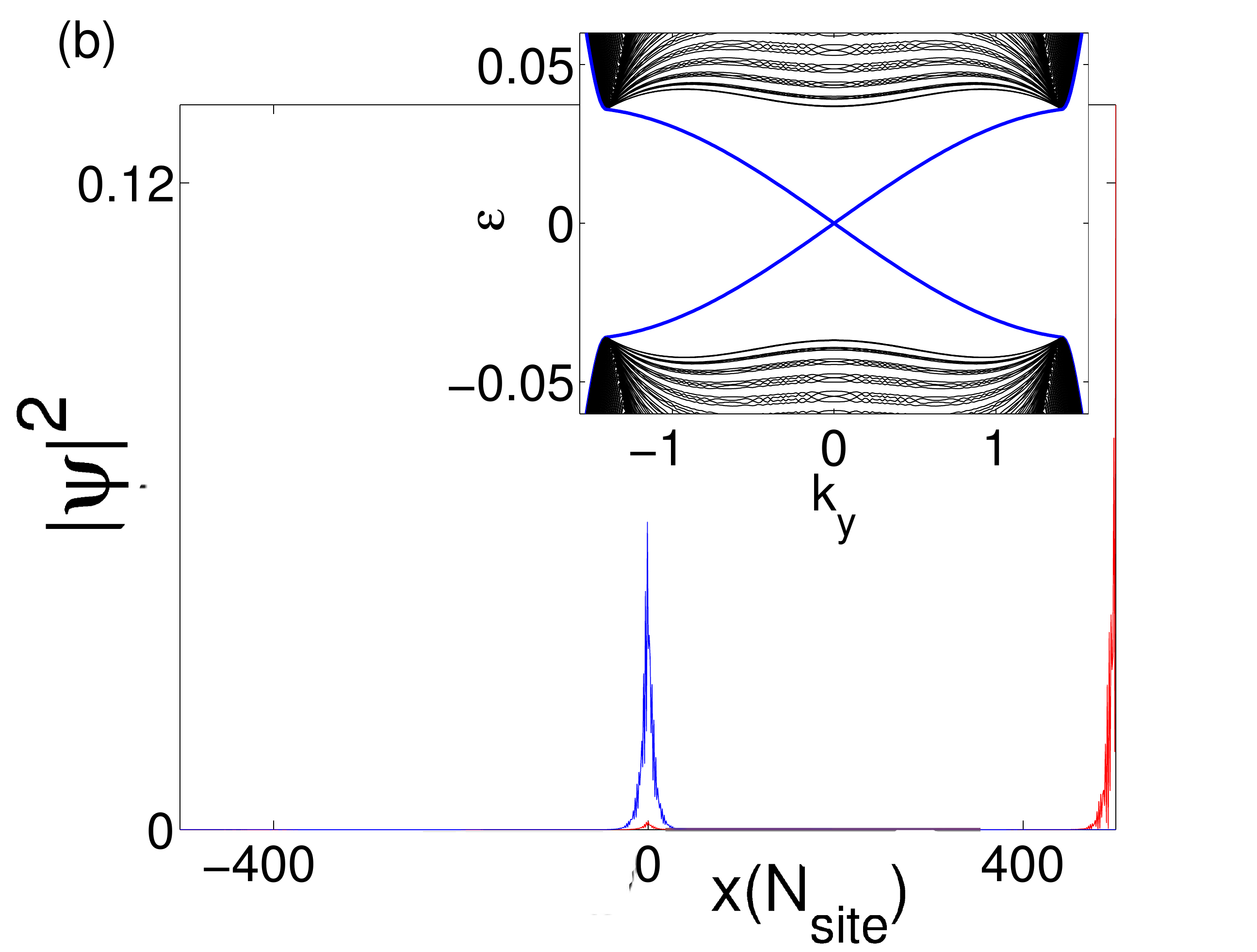}\caption{\textbf{(a) }The Floquet spectrum of the studied three dimensional
model in a domain wall configuration. The dashed red line denotes
modes that are localized at the domain wall, while the blue line denotes
modes that are localized at the edges. The spectrum is doubly degenerate.
Results are for $A=0.2,B=-0.2,M=1,\omega=2.7,V_{0}=0.5,L=300$. Note
that for $L=300$, a small mixing between the edge and domain wall
modes exists. This mixing decays exponentially with the system size.\textbf{
(b)} The amplitude of the localized modes for $L=1000$. The inset
denotes the Floquet spectrum when the radiating light is uniform.
\label{Flo:HgTe3D}}
\end{figure}

In order to demonstrate that on-resonance light can induce topological
properties in this system, we consider linearly polarized radiation,
described by the scalar potential term $V\left(t\right)=V_{0}\gamma_{4}\cos\left(\omega t+\alpha\right)$,
where $V_{0}$, $\alpha$ and $\omega$ are constants. We place the
system in a toroidal geometry, with open boundaries in one direction
and periodic in the remaining two. The inset of Figure \ref{Flo:HgTe3D}(b)
shows that a single Dirac cone now exists at each boundary of the
system, in agreement with Ref. \cite{Lindner3d}. These states are
topologically protected. In particular, we find numerically that the
edge modes are robust against weak breaking of PH symmetry.

We now replace the uniform light by a domain wall configuration in
which the potential changes sign along the $z$ direction. Figure
\ref{Flo:HgTe3D}(a) shows that in addition to the edge modes, the
spectrum now includes a pair of Dirac cones that are confined in the
region near the domain wall.

We explain these results by evaluating the Floquet Hamiltonian. Since
the external radiation is on-resonance, we can use the RWA for this
purpose. As a first step, we omit the components of $V^{\mu}$ that
are parallel to $D_{k}^{\mu}$ by defining $V_{\perp}^{\mu}=V^{\mu}-\frac{V\cdot D}{D^{2}}D^{\mu}=\frac{D_{4}}{D^{2}}V_{0}\left(-\vec{d}_{k},\frac{d_{k}^{2}}{D_{4}}\right)$.
For uniform light, this procedure yields (for $\alpha=0$)
\begin{equation}
H_{F}=\left(1-\frac{\omega}{2D_{k}}\right)D_{k}\cdot\gamma+\frac{1}{2}V_{\perp}\cdot\gamma.\label{eq:General 3D Hf}
\end{equation}
In the low energy limit, $D_{k}^{\mu}\approx\left(A\vec{k},M\right)+O\left(k^{2}\right)$
and Eq. (\ref{eq:General 3D Hf}) becomes

\begin{equation}
H_{F}\approx\left(\left(\mu\frac{A}{M}+\Delta_{0}\right)\vec{k},\eta\right)^{\mu}\gamma_{\mu}\label{eq:Hf 3D}
\end{equation}
where $\mu=\frac{\omega}{2}-M$ and $\Delta_{0}=\frac{V_{0}A}{2M}$
. 

The domain wall configuration can be formulated as a sign flip of
$\Delta_{0}$ at $z=0$, which leads to the BdG equation

\begin{equation}
\left(\mu\frac{A}{M}+\Delta_{0}\left(z\right)\right)\left(\vec{k}_{2}\cdot\vec{\gamma}-i\gamma_{3}\partial_{z}\right)\psi=\left(\varepsilon-\mu\gamma_{4}\right)\psi\label{eq:BdG HgTe3D}
\end{equation}
where $\vec{k}_{2}=\left(k_{x},k_{y}\right)$. Equation (\ref{eq:BdG HgTe3D})
has quasi-stationary solutions, localized at the domain wall. These
states correspond to $k_{x}=k_{y}=0$, for which the EOM reduce to
$\left(-i\left(\mu\frac{A}{M}+\Delta_{0}sign\left(z\right)\right)\partial_{z}+\gamma_{3}\gamma_{4}\mu\right)\psi=0$.
The last equation has four normalizable solutions localized at the
domain wall, provided that $\Delta_{0}>\mu\frac{A}{M}$.

The four quasi-stationary solutions split into two Kramers pairs associated
with time reversal symmetry. In addition, they are eigenstates of
the particle-hole operator, $\gamma^{5}$, such that the two Kramers
pairs transform differently under this transformation. As a result,
the bulk modes are protected against mixing when TRS and PHS are present.
When PH symmetry is weakly broken, a small gap is opened in the spectrum.
For example, for $\varepsilon_{k}=-\varepsilon_{0}\left(\cos k_{x}+\cos k_{y}+\cos k_{z}-3\right)$
, $\varepsilon_{0}=B/2$, the gap is $2.75\times10^{-5}$. The topological
protection of the domain wall modes and the opening of a gap when
PH symmetry is broken is similar in nature to the zincblende bulk
modes discussed in Ref.\cite{Tenebaum Katan  D. Podolsky}.

\section{Summary and Discussion}

In Summary, we have demonstrated that spatially modulated light can
induce dramatic effects in Floquet topological insulators and provided
various schemes to generate these effects. For example, we established
that domain walls in the frequency of the light and domain walls and
vortices in the phase of the light may lead to localized modes with
zero quasi-energy in the bulk of a system described by the studied
zincblende model. In addition, we found that similar effects can be
realized in a honeycomb lattice such as graphene, by domain wall and
vortex configurations of the phase of the light and by an interface
between light beams with different polarizations. We also provided
a generalization of these results to a three dimensional model. Our
work illustrates the great potential and versatility of modulated
FTI as topological phases of matter.

Let us now briefly discuss the physical manifestations of our results
in realistic systems. Currently, the most promising proposal for the
application of modulated FTIs is in the field of optics. Recent work
\cite{Photonic Crystals 2,Photonic Crystal} demonstrated that FTIs
can be realized in engineered photonic systems. The phenomenology
of these setups include protected boundary modes. For example, in
Ref. \cite{Photonic Crystal} helical waveguides were arranged on
a honeycomb lattice in order to create an FTI with a chiral edge mode
and no backscattering. By simple modifications of the waveguide configuration,
it may be possible to implement our results in these experiments.
In this case, the helicity of the light plays the role of circularly
polarized light, and by rotating helices about their screw axis, one
can control the phase of the external perturbation locally. Thus,
for example, by arranging the helices in a vortex configuration it
may be possible to simulate the Caroli-de Gennes-Matricon vortex core
states of a superconductor\cite{Caroli}.

Applications of our results in solid state systems may also be possible,
but some issues must be resolved for this to be accomplished, especially
in cases where the driving frequency is on-resonance. Mainly, in this
case it is difficult to determine what will be the occupation of states
for realistic systems. Early work in this field demonstrated that
in certain special cases, resonantly-driven systems may achieve a
steady state with occupation that is given by a Fermi-Dirac distribution
of the quasi-energies, with the effective temperature determined by
the interaction and phonon relaxation rates \cite{Glazman,Eliashberg,Galitskii}.
 In addition, some experimental work on the occupation of electronic
states in radiated systems also exists \cite{Haug,Goebel,Knox}. An
important result in this context is the direct observation of an energy
gap in a Floquet system \cite{Haug}. Yet, a general understanding
of the occupation of states in systems driven on-resonance is still
lacking. On the other hand, systems driven off-resonance are much
easier to understand and perhaps hold the greatest potential for solid
state realizations. \cite{Fertig Gu Arovas Auerbach,Oka_Aoki,Kitagawa_Oka_Brataas_Fu_Demler}.
For example, the conductivity of these systems when connected to external
leads has been shown to behave as expected for topological phases
with gapless modes \cite{Fertig Gu Arovas Auerbach,Kitagawa_Oka_Brataas_Fu_Demler,kundu}.
Our results suggest that it is possible to induce anisotropic conductivity
in these systems by using space modulated light.

\section{Acknowledgments}

We would like to thank Adi Stern for the suggestion to consider frequency
modulation. This research was supported by the Israel Science Foundation
and by the E.U. under grant agreement No. 276923 \textendash{} MC--MOTIPROX.


\begin{thebibliography}{References}
\bibitem{Fu_Kane}L. Fu and C.L. Kane, Phys. Rev. Lett. \textbf{100},
096407 (2008).

\bibitem{Qi Zhang} X. L. Qi, S.C. Zhang, Rev. Mod. Phys. \textbf{83},
1057 (2011).

\bibitem{Hassan_Kane}M.Z. Hassan and C.L. Kane, Rev. of Mod. Phys.,
\textbf{82} (2010).

\bibitem{KaneMele}C. L. Kane and E. J. Mele, Phys. Rev. Lett. \textbf{95},
146802 (2005).

\bibitem{Bernevig}B.A. Bernevig, T. L. Hughes, and S.C. Zhang, Science
\textbf{314}, 1757 (2006). 

\bibitem{Kitagawa_Berg_Demler_Rudner_033429}T. Kitagawa, M. S. Rudner,
E. Berg, and E. Demler, Phys. Rev. A \textbf{82}, 033429 (2010). 

\bibitem{Konig Wiedrmann}M. König, S. Wiedmann, C. Brüne, A. Roth,
H. Buhmann, L. W. Molenkamp, X. L. Qi, and S. C. Zhang, Science \textbf{318}
(5851), 766 (2007).

\bibitem{Xia Qian Hsieh}Y. Xia, D. Qian, D. Hsieh, L. Wray, A. Pal,
H. Lin, A. Bansil, D. Grauer, Y. S. Hor, R. J. Cava, and M. Z. Hasan,
Nat. Phys. \textbf{5}, 398 (2009).

\bibitem{Hsieh Qian}D. Hsieh, D. Qian, L. Wray, Y. Xia, Y. S. Hor,
R. J. Cava, and M. Z. Hasan, Nat. Phys. \textbf{452}, 970 (2008).

\bibitem{Kitagawa_Berg_Demler_Rudner_235114}T. Kitagawa, E. Berg,
M. Rudner, and E. Demler, Phys. Rev. B \textbf{82}, 235114 (2010).

\bibitem{LRG}N.H. Lindner, G. Refael, V. Galitski, Nat. Phys. 7,
490 (2011).

\bibitem{Lindner3d}N.H. Lindner, D. L. Bergman, G. Refael, V. Galitski,
Phys. Rev. B \textbf{87}, 235131 (2013).

\bibitem{Jiang}L. Jiang, T. Kitagawa, J. Alicea, A. R. Akhmerov,
D. Pekker, G. Refael, J. I. Cirac, E. Demler, M. D. Lukin, and P.
Zoller, Phys. Rev. Lett. \textbf{106}, 220402 (2011).

\bibitem{Kitagawa_Oka_Brataas_Fu_Demler}T. Kitagawa, T. Oka, A. Brataas,
L. Fu, and E. Demler, Phys. Rev. B \textbf{84}, 235108 (2011).

\bibitem{Fertig Gu Arovas Auerbach}Z. Gu, H.A. Fertig, D. P. Arovas,
and A. Auerbach, Phys. Rev. Lett. \textbf{107}, 216601 (2011).

\bibitem{Photonic Crystal}M. C. Rechtsman, J. M. Zeuner, Y. Plotnik,
D. Podolsky, F. Dreisow, Y. Lumer, S. Nolte, M. Segev, A. Szameit,
Nature \textbf{496}, 196 (2013).

\bibitem{Photonic Crystals 2}T. Kitagawa, M. A. Broome, A. Fedrizzi,
M. S. Rudner, E. Berg, I. Kassal, A. G. White, A. Aspuru-Guzik, E.
Demler, Nature Comm. \textbf{3}, 882 (2012). 

\bibitem{Tenebaum Katan  D. Podolsky}Y. Tenenbaum Katan, D. Podolsky,
Phys. Rev. Lett. \textbf{110}, 016802 (2013).

\bibitem{Thouless}D. J. Thouless, M. Kohmoto, M. P. Nightingale,
and M. den Nijs, Phys. Rev. Lett. \textbf{49}, 405 (1982). 

\bibitem{Eastham}M. S. P. Eastham, \textit{The spectral theory of
periodic differential equations}, Scottish Academic Press, (1973).

\bibitem{Rudner} M. S. Rudner, N. H. Lindner, E. Berg, M. Levin,
Phys. Rev. X 3, 031005 (2013).

\bibitem{Schrieffer}J.R. Schrieffer, \textit{Theory of Superconductivity}
(Addison-Wesley, Reading, MA), (1964).

\bibitem{Read &Green}N. Read and D. Green. Phys. Rev. B \textbf{61},
10267 (2000).

\bibitem{TeoKane} J. C.Y. Teo and C. L. Kane, Phys. Rev. B \textbf{82},
115120 (2010).

\bibitem{Castro Neto} A. H. Castro Neto, F. Guinea, N.M.R Peres,
K.S. Novoselov and A.K. Geim, Rev. of Mod. Physics, \textbf{81}, 109
(2009).

\bibitem{Torres2}Hernan L. Calvo, Horacio M. Pastawski, Stephan Roche,
Luis E. F. Foa Torres Appl. Phys. Lett. \textbf{98}, 232103 (2011).

\bibitem{Torres}P. M. Perez-Piskunow, G. Usaj, C. A. Balseiro, and
L. E. F. Foa Torres, arxiv:1308.4362 .

\bibitem{Roslyak_Gumb_Huang}O. Roslyak, G. Gumbs, and D. Huang, J.
Appl. Phys. \textbf{109}, 113721 (2011).

\bibitem{Oka_Aoki}T. Oka and H. Aoki, Phys. Rev. B \textbf{79}, 081406
(2009).

\bibitem{Caroli}C. Caroli, P. G. de Gennes,J. Matricon, Phys. Rev.
Lett. \textbf{9}, (1964).

\bibitem{Josephson in P-wave}H.J. Kwon, K. Sengupta, V. M. Yakovenko,
The European Physical Journal B \textbf{37}, 349-361 (2004).

\bibitem{Schneider}A. P. Schnyder, S. Ryu, A. Furusaki, A. W. W.
Ludwig ,Phys. Rev. B \textbf{78}, 195125 (2008).

\bibitem{Eliashberg}G. M. Eliashberg, JETP Lett. \textbf{11}, 114
(1970).

\bibitem{Glazman}L.I. Glazman, Sov. Phys. JETP \textbf{53}, 178 (1981). 

\bibitem{Galitskii}V. M. Galitskii, S. Goreslavskii, and V. F. Elesin,
Sov. Phys. JETP \textbf{30}, 117 (1970). 

\bibitem{Haug}Q. T. Vu, H. Haug, O. D. Mucke, T. Tritschler, M. Wegener,
G. Khitrova, and H. M. Gibbs, Phys. Rev. Lett. \textbf{92}, 217403
(2004). 

\bibitem{Goebel}E. O. Goebel and O. Hildebrand, Phys. Status Solidi
B \textbf{88}, 645 (1978). 

\bibitem{Knox}W. H. Knox, D. S. Chemla, G. Livescu, J. E. Cunningham,
and J. E. Henry, Phys. Rev. Lett.\textbf{ 61}, 1290 (1988). 

\bibitem{kundu}A. Kundu, B. Seradjeh, arXiv:1301.4433 (2013).\end{thebibliography}
\end{document}